\documentclass[pre,preprint]{revtex4-1}
\usepackage{graphicx,xcolor}
\usepackage[colorlinks,linkcolor=blue,citecolor=blue]{hyperref}
\usepackage{multirow}
\begin{document}

%%%%%%%%%%%%%%%%%%%%%%%%%%%%%%%%%%%%%%%%%%%%%%%%%%%%%%%%%%%%%%%%%%%%%%%%%%%%%%%%%%%%%%%%%%%%%%%%%%%%%%%%%%%%%%%%%%%%%%%%%%%%%%%%%%%%%%%%%%%%
\title{Experimental investigation of vortex ring evolution in polymer solution}

\author{Swastik Hegde}
\affiliation{Engineering Mechanics Unit, Jawaharlal Nehru Centre for Advanced Scientific Research,
Bangalore 560064, India}
\author{Shashank H J}
\affiliation{Engineering Mechanics Unit, Jawaharlal Nehru Centre for Advanced Scientific Research,
Bangalore 560064, India}
\author{K R Sreenivas}
\affiliation{Engineering Mechanics Unit, Jawaharlal Nehru Centre for Advanced Scientific Research,
Bangalore 560064, India}
%\affiliation{Engineering Mechanics Unit, Jawaharlal Nehru Centre for
%Advanced Scientific Research, \\Jakkur PO, Bangalore 560064, India
%}

%%%%%%%%%%%%%%%%%%%%%%%%%%%%%%%%%%%%%%%%%%%%%%%%%%%%%%%%%%%%%%%%%%%%%%%%%%%%%%%%%%%%%%%%%%%%%%%%%%%%%%%%%%%%%%%%%%%%%%%%%%%%%%%%%%%%%%%%%%%%

\date{\today}

%%%%%%%%%%%%%%%%%%%%%%%%%%%%%%%%%%%%%%%%%%%%%%%%%%%%%%%%%%%%%%%%%%%%%%%%%%%%%%%%%%%%%%%%%%%%%%%%%%%%%%%%%%%%%%%%%%%%%%%%%%%%%%%%%%%%%%%%%%%%

\begin{abstract}
We have conducted experiments on the effect of polymer solutions on the formation and propagation of vortex rings. We study this effect in aqueous solution of hydrolyzed polyacrylamide (PAMH) at different concentrations. Addition of PAMH imparts shear-rate dependent viscosity and elasticity to the solvent. With increasing concentration of PAMH both the zero-shear-rate viscosity ($\eta_0$) and the infinite-shear-rate viscosity ($\eta_{\infty}$) increase. The relaxation time also increase with the increase in concentration. We generate vortex rings using a piston-cylinder mechanism in a glass tank and measure vortex ring properties such as ring position, ring circulation, enstrophy, kinetic energy and peak vorticity using particle image velocimetry (PIV). Experiments are conducted by (1) matching impulse, and (2) matching Reynolds number. We show that, at constant impulse, vortex ring properties in PAMH solution deviate from that in Newtonian water. As the concentration of the PAMH solution increases, so too does the deviation from water. We further perform experiments in water by matching the limiting Reynolds numbers of the PAMH solution corresponding to both $\eta_0$ and $\eta_{\infty}$. We find that while the circulation of PAMH solutions lie between the circulation curves of the two extreme Reynolds number matched water experiments for an extended period of time, the enstrophy and peak vorticity do not. We attribute this behaviour to the modification of vorticity distribution within the core of vortex rings in PAMH solutions. We also study the effect of polymer solutions on the formation number. We find that the formation number remains the same even in polymer solutions. We also demonstrate, using planar laser induced fluorescence (PLIF), the phenomenon of ring reversal. Once the vortex ring stops, it begins to unwind and retract by translating and rotating in the opposite direction. We attribute this behaviour of ring reversal to the elastic properties of polymer solutions.
\end{abstract}
%%%%%%%%%%%%%%%%%%%%%%%%%%%%%%%%%%%%%%%%%%%%%%%%%%%%%%%%%%%%%%%%%%%%%%%%%%%%%%%%%%%%%%%%%%%%%%%%%%%%%%%%%%%%%%%%%%%%%%%%%%%%%%%%%%%%%%%%%%%%
\maketitle

%\documentclass[final,5p,twocolumn]{elsarticle}
%\documentclass{article}
%%%%%%%%%%%%%%%%%%%%%%%%%%%%%%%%%%%%%%%%%%%%%%%%%%%%%%%%%%%%%%%%%%%%%%%%%%%%%%%%%%%%%%%%%%%%%%%%%%%%%%%%%%%%%%%%%%%%%%%%%%%%%%%%%%%%%%%%%%%%

%\linenumbers
%%%%%%%%%%%%%%%%%%%%%%%%%%%%%%%%%%%%%%%%%%%%%%%%%%%%%%%%%%%%%%%%%%%%%%%%%%%%%%%%%%%%%%%%%%%%%%%%%%%%%%%%%%%%%%%%%%%%%%%%%%%%%%%%%%%%%%%%%%%%
% \maketitle
%%%%%%%%%%%%%%%%%%%%%%%%%%%%%%%%%%%%%%%%%%%%%%%%%%%%%%%%%%%%%%%%%%%%%%%%%%%%%%%%%%%%%%%%%%%%%%%%%%%%%%%%%%%%%%%%%%%%%%%%%%%%%%%%%%%%%%%%%%%%
\section{Introduction}

Vortex rings are self-driven coherent structures and vortices are ubiquities in turbulent flows. Vortex rings are studied extensively for their rich physics and their relevance in various flow contexts. Their interaction with density interface is relevant in understanding the mixing process \cite{linden1973interaction, advaith2017interaction}, vortex-breakdown for understanding the transition to turbulence, quantum turbulence \cite{kerr2010numerical} and entrainment process in turbulent jets and plumes \cite{liepmann1992role, sreenivas2000vortex}.  Interaction of vortices and their dynamics plays an important role in energy cascade in turbulent flows.  Gharib et. al. \cite{gharib2006optimal} have proposed a method to measure cardiac health by determining the strength of vortex rings formed as blood passes through the valves within the heart, and recently it has been shown that vortex rings can be used for cell encapsulation \cite{an2016mass}. These biological fluids are known to exhibit non-Newtonian behavior, as do fluids in many industrial flows. The study of vortex rings in non-Newtonian fluids is thus very relevant. However, there are only a few detailed studies on vortex rings in non-Newtonian fluids. \\

%The vortex ring is a fundamental and fascinating flow structure. Systematic studies of vortex rings date back to Reynolds's experiments in 1876 \cite{reynolds1876resistance}. Saffman, in his famous book on vortex dynamics \cite{saffman},  describes vortex rings as follows: ``One particular motion exemplifies the whole range of problems of vortex motion and is also a commonly known phenomenon, namely, the vortex ring .... Their formation is a problem of vortex sheet dynamics, the steady state is a problem of existence, their duration is a problem of stability, and if there are several we have a problem of vortex interactions.'' Vortex rings are ubiquitous in numerous real world flows. The understanding of the behaviour of vortex rings has proven to be useful in applications ranging from aircraft design to the propulsion mechanism of squids and jellyfish. There is a growing trend of vortex rings being applicable in biological flows - Gharib et. al. \cite{gharib2006optimal} have proposed a method to measure cardiac health by determining the strength of vortex rings formed as blood passes through the valves within the heart; recent studies show that vortex rings can be used for cell encapsulation \cite{an2016mass}. These biological fluids are known to exhibit non-Newtonian behavior, as do fluids in many industrial flows. The study of vortex rings in non-Newtonian fluids is thus very relevant in today's world. However, there is a dearth of detailed studies on vortex rings in non-Newtonian fluids. \\

Palacious-Morales and Roberto Zenit \cite{palacios2013formation} have studied the behaviour of the vortex ring in shear-thinning liquids at low Reynolds number. The shear-thinning fluids have been characterized using the power-law model. The Reynolds number of the shear-thinning vortex ring was calculated using a characteristic shear rate, and the ring thus generated was compared with a Newtonian vortex ring at the same Reynolds number. They find that the shear-thinning fluid retards the vortex ring, leading to a reduction in its propagation properties. Bentata et al. \cite{bentata2018experimental} studied vortex rings in shear-thinning solutions at low Reynolds number ($Re \approx$ 30 to 300). Their findings of the propagation properties of the vortex ring, for instance the ring velocity and ring diameter are in agreement with that obtained by Palacios \cite{palacios2013formation}. In another study, Palacios-Morales et. al. \cite{palacios2015negative} studied the effect of viscoelasticity on vortex rings at low $Re$. The vortex ring showed a reduction in its circulation at the end of the formation stage. Formation of a `negative' vortex ring ahead of the primary ring was also observed.\\

A shear-thinning fluid typically exhibits viscosity plateaus at both zero-shear-rate ($\eta_0$) and infinite-shear-rate ($\eta_{\infty}$). The Carreau-Yasuda model, which captures these plateaus, is thus better suited for characterization of these polymer solutions than the power-law model \cite{olsthoorn2014dynamics}. Depending on the shear rates, vortex ring can experience two distinctive viscosity variation, (1) power-law variation, and (2) a constant viscosity corresponding to the viscosity plateaus. The reduction in the shear rate as the ring propagates forward leads to an increase in the local viscosity, which further retards the ring. Crucially, at low $Re$ (as seen in the experiments of Palacios-Morales \cite{palacios2013formation} and Bentata et. al. \cite{bentata2018experimental}), the shear rates experienced by the ring is at the low shear rate end of power-law region, and quickly enter into the zero shear plateau as the ring propagates forward. Additionally, in the aforementioned studies, the viscosity of the non-Newtonian fluids is significantly larger than the water viscosity.\\
 
% In case of vortex rings at such low $Re$, shear-rate over most part of the field quickly falls below characteristic shear-rate. Hence, it is not surprising to see that newtonian ring does better than shear-thinning vortex rings. Hence, approximating whole field's viscosity variation by power law would be erroneous. Also there is an inherent limitation in matching Reynolds number when a shear thinning fluid is involved, which is one of the findings of our work. \\

%  It is very important to note here that, for such low inertia vortex rings, viscosity experienced by the ring starts from lower shear-rate end of power law region, and as the ring translates forward whole flow field viscosity very quickly reaches zero shear viscosity plateau. This would explain the main observation of Bentata et al. \cite{bentata2013etude}, i.e. power law affecting formation of the ring, and translation of the ring being nearly independent of the power law index. However, this is not complete, because when vortex rings are generated with high inertia, vortex ring would initially experience infinite shear viscosity. As the ring slows down, would then experience power law viscosity regime, and finally gets dissipated by viscosity corresponding to zero shear rate. This scenario would then mean that, formation stage is independent of power law index, whereas translation of the ring is dependent on the power law index. Hence, in order to cover bigger picture, vortex rings with higher inertia has to be studied. We have carried out experiments to fill in the required gap. \\

We study vortex rings in non-Newtonian polymer solutions at large Reynolds numbers. We choose polymer solution viscosities that are akin to those used in polymer drag reduction applications. The shear viscosity of the non-Newtonian fluids in this study are of the order of magnitude of Newtonian water. PAMH solutions used in this work are shear thinning, and are characterized using the Carreau-Yasuda model. The shear rates corresponding to the Reynolds numbers considered are closer to the infinite-shear-rate plateau. Thus the vortex ring experiences the whole power-law region and the zero shear plateau, as opposed to the shear rates observed in literature. This study of the canonical form of vorticity in the presence of polymer solutions helps in building toy models to shed light on the mechanism of the complex phenomenon of turbulent polymer drag reduction.\\

In this paper, we experimentally compare vortex ring formation and its free-shear motion in PAMH polymer solution with that in water. Key quantities of vortex-ring evolution, such as circulation, enstrophy, kinetic energy and peak vorticity are measured using PIV. We report the key difference regarding vorticity distribution in two scenarios observed in our experiments. We also demonstrate that the Reynolds number matching alone for vortex rings in two different solutions would not account for their behaviour. We also show that formation number is independent of the fluid properties used in the current experiments.

\section{Experiments}

\begin{figure*}[h]
	\centering
	\includegraphics[width=\textwidth]{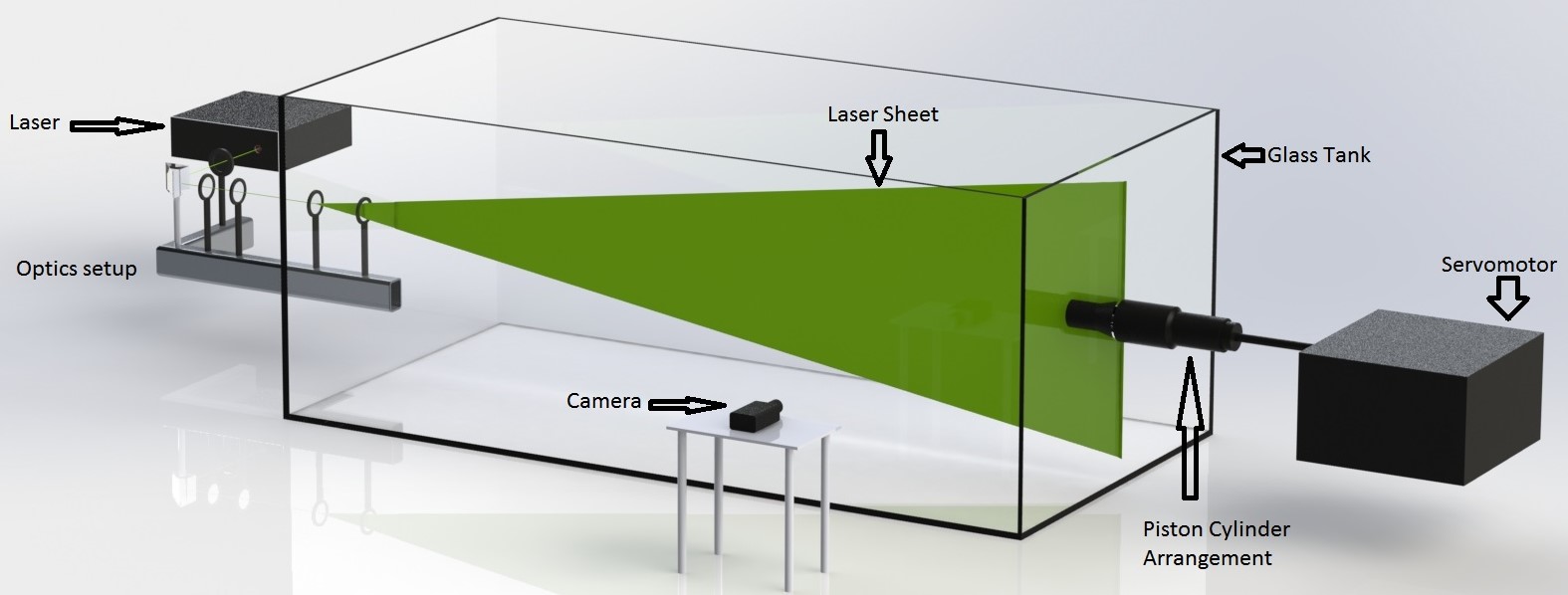}
	\caption{Schematic of experimental setup}
	\label{fig: Schematics of Experimental setup}
\end{figure*}

We generate vortex rings using a piston-cylinder mechanism driven by a servomotor, with control over the piston speed ($V$) and the piston stroke length ($L$). The servomotor operates on a closed-loop servo system, with an optical encoder. We generate the vortex rings in a glass tank (3 $m$ $\times$ 0.75 $m$ $\times$ 0.75 $m$) with a capacity of $\approx$ 1500 litres. A nozzle with exit diameter ($D$) of 50 $mm$ is connected at the end of the piston-cylinder arrangement. The exit of the nozzle is located at a distance of 300 $mm$ ($6D$) from both the bottom of the tank and the free surface of the fluid. The exit of the nozzle is 300 $mm$ from the end of the back wall of the tank and 300 $mm$ from the side walls of the tank. The walls of the tank are sufficiently far away from the nozzle to make end effects negligible. A schematic of the experimental arrangement is shown in Fig. \ref{fig: Schematics of Experimental setup}. \\    

% A schematic of the experimental setup is shown in figure \ref{fig: Schematics of Experimental setup}. Experiments are conducted in glass tank which is three meter in length and one meter in width, with a capacity of 1200 litre. Tank is filled with the solution in which vortex rings are to be studied. Vortex rings are generated using piton cylinder mechanism driven by a servomotor. A computer program controls the speed (V) and stroke-length (L) of the piston through servomotor. A nozzle with exit diameter ($\mathrm{D}$) of 50 $mm$ is connected at the end of the piston cylinder arrangement. Nozzle end is located at a distance of 300 $mm$ ($\mathrm{6D}$) from both bottom and backside wall, and at the same distance from free surface of the liquid. \\

We perform 2-D particle image velocimetry (PIV) and planar laser induced fluorescence (PLIF) to study the vortex rings. A vertical laser sheet is generated from one end of the tank to the other using a combination of spherical and cylindrical lenses. We utilise a Litron (nano-series) pulsed, dual-head Nd-YAG laser, with a wavelength of 532 $nm$. The lasers are operated at an intensity of 100 $mJ$ per pulse with a repetition rate of 25 $Hz$. The time delay between the two laser heads is controlled using a Stanford delay generator. We utilise a Phantom Speedsense 9040 high speed camera with a resolution of 2 megapixels to acquire the image maps. The camera is synchronized with the laser using the same Stanford delay generator. We use a $\Delta T$ of 1 $ms$ between the two laser heads to ensure that the strong vortical regions of the vortex rings are adequately captured. These strong vortical regions require a good concentration of seeding particles to obtain good correlations in the PIV analysis. We add 1 $g$ of seeding particles (polyamide particles of size 50 $\mu m$ and density 1.2 $g/cm^3$) exclusively to vortex rings over the trials, in addition to 2 $g$ of seeding particles that are added to the whole tank. The PIV analysis is performed using an interrogation window of 32 pixels $\times$ 32 pixels with an overlap of 50\%. Further information about PIV can be found in \cite{adrian2005twenty, adrian2011particle, prasad2000particle,raffel2018particle,westerweel1997fundamentals}. Our PIV analysis yields a 2-D velocity field, with radial and axial components of velocities. Additional details of the experimental arrangement can be found in \cite{SwastikHegde}.

% One end of the tank houses the vortex ring generating mechanism, and the laser sheet enters from the opposite end, passing through the central plane of the nozzle vertically. 100 $mJ$ Nd:YAG pulsed laser (Litron Nano-series) operating at 532 $nm$ with 25 $Hz$ frequency is used for PIV experiments. The laser beam coming out of the laser is converted to a laser sheet using a combination of lenses. The spread of the laser sheet is such that, nearly parallel rays enter the area of interest. \\

% Phantom v9.1 high speed camera with 1632 $\times$ 1200 pixels (2 MP) is used for image acquiring. Images are acquired at 25 $Hz$, and the camera is synchronized with the Laser using Stanford delay generator. Images are captured with 8-bit depth. Exposure time of the camera is kept at 900 $\mu s$ so that, two Laser shots which are separated by a $dt$ of 1 $ms$ won't overlap. 17 -35 $mm$ lens is used for the camera magnification. 50 micron polyamide particles with density of 1.2 $g/cm^3$ are used as tracers. 2 $g$ of particles are added in the whole tank. Additional 1 $g$ of particles is added to vortex rings over the trials to get better correlation inside the ring, which is necessary since the vortex ring is a highly vortical region. Details of the experimental setup can be found in \cite{SwastikHegde}.

% \section{Methodology}

In this work, we extract the properties of vortex ring as it propels forward due to self-induced velocity. We calculate the peak azimuthal ring vorticity, enstrophy, kinetic energy, circulation and ring position. We evaluate the azimuthal vorticity by taking a curl of the 2-D velocity vectors. We calculate the enstrophy and kinetic energy over the whole field. The ring circulation is evaluated over one half of the ring. Finally, we evaluate the ring position by finding the locations corresponding to the peak azimuthal vorticities within the ring.\\ 

% Interrogation window size of 32 $\times$ 32 pixel with 50 $\%$ overlap is used for PIV measurements. PIV gives 2D velocity field, with radial and horizontal components of velocity. Azimuthal vorticity is calculated using this velocity data. Details of particle image velocimetry can be found in \cite{adrian2005twenty, adrian2011particle, prasad2000particle,raffel2018particle,westerweel1997fundamentals}. Peak vorticity location corresponds to the centre of core of the ring, and is traced to get distance traversed by vortex ring \cite{querzoli2010flow,gharib1998universal}. Enstrophy and kinetic energy are calculated over the whole field. However, vorticity field is integrated over one half of the ring to calculate circulation, since the other half will have the same circulation with opposite sign, and would cancel-out if calculated over whole domain. \\

We study the effect of polymer solutions on vortex rings by performing four types of experiments: (1) matched impulse experiments (Section \ref{section: Matched Impulse experiments}), (2) matched $Re$ (Section \ref{section: Re matched experiments}) experiments, (3) formation number ($L/D$) experiments (Section \ref{section: formation number}), and (4) vortex ring reversal (Section \ref{section: elastic effects}). For (1) and (2), the piston stroke-length $L$ is kept constant at 100 $mm$, and piston velocity $V$ is varied. With this arrangement, we generate vortex rings over a large range of $Re$ (1000 to 100000). The Reynolds number for the current experiments is calculated based on the piston input parameters - $Re = L V/\nu$. For (3), the piston velocity $V$ is maintained constant at 500 $mm/s$, while the stroke length $L$ is varied to obtain different values of $L/D$. For (4), PLIF is performed at low values of $Re$. The nomenclature used in this paper is Lxxx Vyyy, where xxx and yyy corresponds to the stroke length $L$ and piston velocity $V$, respectively. For instance, L100 V500 corresponds to an experiment with a stroke length $L$ of 100 $mm$ and a piston velocity $V$ of 500 $mm/s$.\\

The polymer solution is prepared in de-ionised (DI) water with a resistivity of 4 M$\Omega m$. The required quantity of polymer is mixed in a container (of about one-tenth the volume of the tank) using a mechanical stirrer, yielding a thickened solution. This thickened solution is then diluted to the required concentration by slowly mixing it with de-ionised water in the glass tank. The solution in the tank is allowed to homogenize for 48 hours, after which the experiments are conducted.\\

% Polymer solution was prepared in DI water with resistivity of 4 M$\Omega$. Solution was sufficiently stirred and kept for 48 hours to homogenize, before the experiments were conducted.

\section{Polymer solution characterization}
We use hydrolyzed polyacrylamide (PAMH) as the polymer in our experiments. The polymer has a molecular weight 12-16 $\times$ $10^6$. We prepare three different polymer solutions with concentrations of 5 $ppm$, 10 $ppm$ and 25 $ppm$. In this section, we measure the rheological properties of these polymer solutions. The shear thinning property of polymer solutions is characterized using the Carreau-Yasuda model. The rheological properties of interest are the zero-shear viscosity $\eta_0$, the infinite-shear viscosity $\eta_{\infty}$, the power-law index $n$, and relaxation time $\lambda$. The first three parameters characterize shear-thinning behaviour, while the relaxation time characterizes the elasticity of the solution.\\ 

\begin{figure*}[h]
	\centering
	\includegraphics[width=\textwidth]{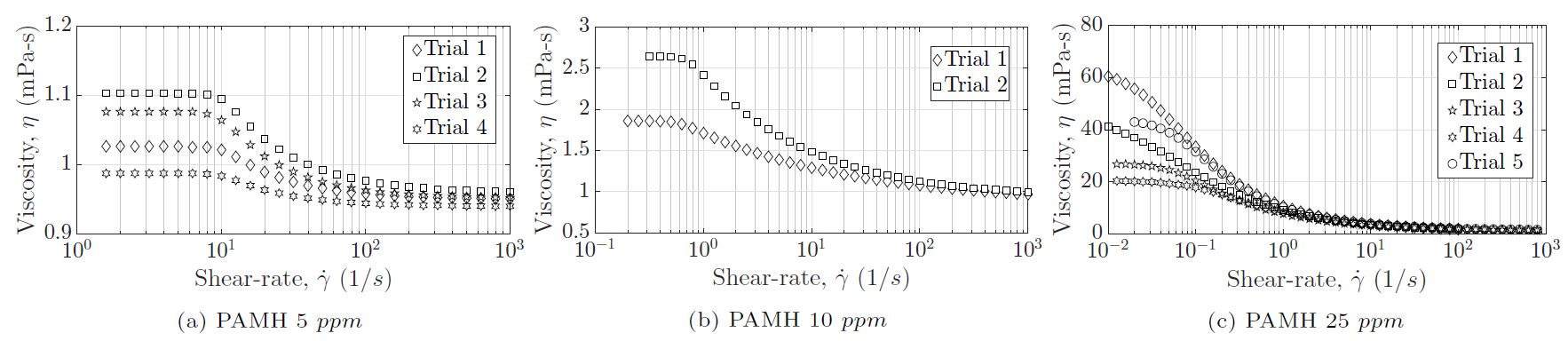}
	\caption{Viscosity vs. shear rate for the PAMH solutions}
	\label{fig: shear thinning}
\end{figure*}

Fig. \ref{fig: shear thinning} shows the shear-thinning viscosity profile measured using rotational rheometry. We use an Anton-Paar MCR 302 rheometer with a double gap geometry to perform these measurements. We see from the plots that, as the shear rate increases, polymers gets stretched and aligned along the flow direction. This leads to a reduced resistance to the flow, thereby leading to a reduced viscosity at high shear-rates. We see from the plots that, the viscosities of the polymer solution increases with the increase in the polymer concentration. This is apparent from the viscosity plots shown for three polymer solutions in Fig. 2a (PAMH 5 $ppm$), Fig. 2b (PAMH 10 $ppm$) and Fig. 2c (PAMH 25 $ppm$). All the rheological properties obtained from Carreau-Yasuda model are shown in Table \ref{table: Properties of polymer solutions}.\\

In our experiments, the shear rate of the shear layer pushed out of the piston, which rolls up to form the vortex ring is $\mathcal{O}(100)$ $1/s$. This means that, during the formation stage of the vortex ring, the viscosity in the shear layer is $\approx$ $\eta_{\infty}$ (see Fig. \ref{fig: shear thinning}). As the vortex ring propagates, the slow reduction in its velocity due to viscous effects causes an increase in the viscosity. This increase in viscosity will reduce the ring velocity further, which leads to a further increase in viscosity, and so on. This process continues till the zero shear plateau is reached, at which the viscosity is $\eta_0$. Therefore, the viscosities experienced by the vortex ring vary from $\eta_{\infty}$ to $\eta_0$ through the power law region. However, the ambient, being stationary, is always at $\eta_0$. The effects of shear thinning are thus significant, and cannot be ignored.\\

We tried to measure the elasticity through oscillatory rheometry using various techniques. However, both storage and loss modulus were very small for the direct measurement. However, in Carreau-Yasuda model, inverse of the shear rate at which the transition from zero-shear plateau to power law region occurs, is a measure of the relaxation time \cite{srinivas2017effect,zhang2018unsteady}. Hence, we obtain the relaxation time from the shear-thinning viscosity profile by using the Carreau-Yasuda model. 
 
% The effect of shear thinning on drag reduction can't be neglected as it is usually done in tubulent drag-reduced flows of shear-thinning fluids \cite{samanta2013elasto}. In fact, shear-thinning viscosity is proposed to be the reason for drag reduction in some studies \cite{PhysRevE.70.055301,PhysRevLett.59.2059,PhysRevLett.92.244503}. 

\begin{table*}[h]
	\begin{center}
		\centering
		\begin{tabular}{|c|c|c|c|c|} % <-- Alignments: 1st column left, 2nd middle and 3rd right, with vertical lines in between
			\hline
			\textbf{Solution} & \textbf{$\lambda$ ($s$)} & \textbf{$n$} & \textbf{$\eta_0$ ($mPa-s$)} & \textbf{$\eta_{\infty}$ ($mPa-s$)}\\
			\hline
			PAMH 5 $ppm$ & 0.1 & \textendash & 1.0 - 1.1 & 0.95\\
			PAMH 10 $ppm$ & 1 - 2 & 0.58 - 0.68 & 2 - 3 & 0.955\\
			PAMH 25 $ppm$ & 12 - 25 & 0.46 - 0.56 & 20 - 60 & 0.96\\
			\hline
		\end{tabular}
		\caption{Carreau-Yasuda parameters for the PAMH solutions}
		\label{table: Properties of polymer solutions}
	\end{center}
\end{table*}

% Efforts to measure the elasticity through oscillatory rheometry was done in great detail. Both storage and loss modulus were very small to be measured. Additional efforts were done by using four different tools (Cone and Plate, Parallel plate, couette and Double gap). Many trials were carried out with reduced gap width in parallel plate tool to measure the elastic properties. However, both storage and loss modulus were too small to be measured. \\

% Hence, for our current study key contribution will be by shear thinning effects. Nonetheless, during final dissipation stage of vortex ring, as the inertia effects are diminished, elastic effects show there presence, which is explained in section \ref{section: elastic effects}.

\section{Vortex Rings at Matched Impulse}
\label{section: Matched Impulse experiments}
In this section, we present experiments on vortex rings in water and in PAMH solutions at the same impulse. Vortex ring experiments where initial impulse provided for polymer solution is same as its Newtonian counterpart is the simplest and rudimentary way to examine the effect of polymers. Experiments at matched impulse can be considered as an analog for certain commercial situations, such as pipeline and pumping
network, where the power input has already been fixed. The impulse is matched by maintaining the same piston parameters, i.e., the stroke length and piston velocity across the different fluids. The piston impulse range is divided into three categories: (1) low impulse - L100 V100, (2) medium impulse - L100 V500, and (3) high impulse - L100 V1000.\\

% Vortex ring experiments where initial impulse provided for polymer solution is same as its Newtonian counterparts is missing in the existing literatures. It is important in certain commercial situations, such as pipeline and pumping network, where the power input has already been at a fixed value. Also, it is most rudimentary, without involving complexities of $Re$ matching, and associated drawbacks (Refer section \ref{section: Re matched experiments} for short-comings of matching $Re$). Same impulse here refers to same piston input i.e. same stroke-length and velocity of piston movement. \\

% The piston impulse range is divided into three categories: (1) low impulse - L100 V100 (2) medium impulse - L100 V500 and (3) high impulse - L100 V1000. Reynolds number calculated based on piston input parameters, $Re = \frac{\Gamma}{\nu} = \frac{\mathrm{L} \times \mathrm{V}}{\nu}$, will be 10000, 50000 and 100000 respectively. Various properties measured using PIV are presented below. \\

\begin{figure*}[h]
	\centering
	\includegraphics[width=\textwidth]{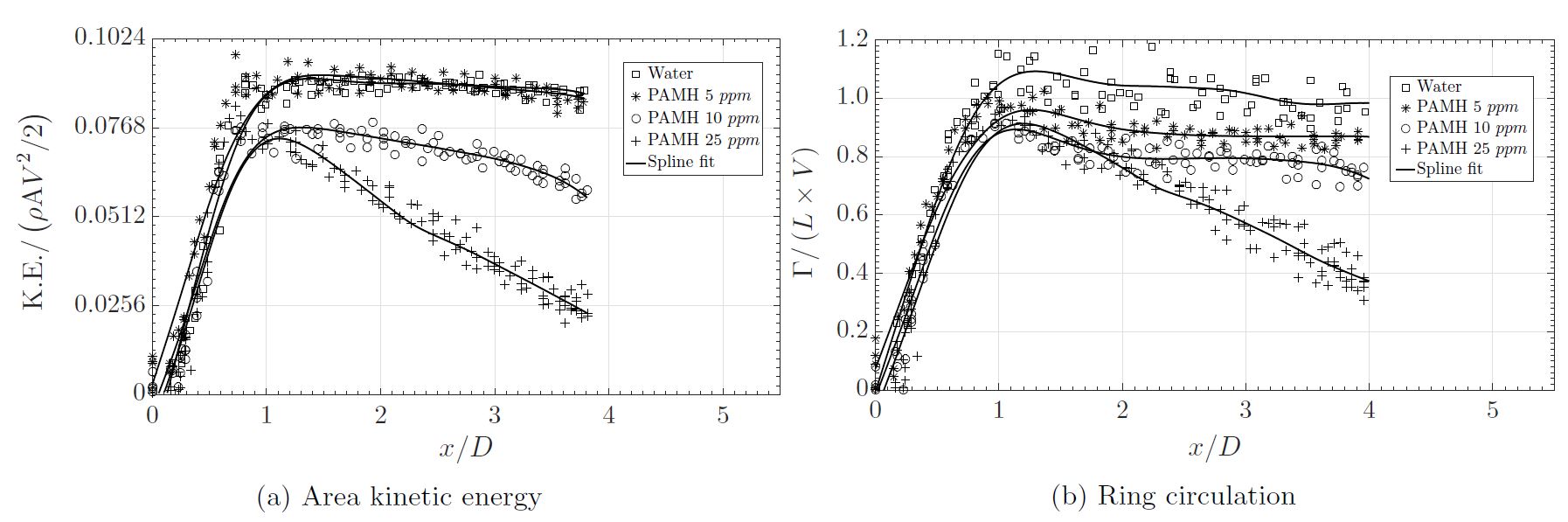}
	\caption{Area kinetic energy and ring circulation for all fluids at L100 V500.}
	\label{fig: Circulation KE constant impulse}
\end{figure*} 

Fig. \ref{fig: Circulation KE constant impulse} shows the behaviour of the non-dimensional ring circulation and the non-dimensional area kinetic energy for all fluids at L100 V500. The evolution of ring circulation is shown in Fig. 3b. We see from the plots that the ring circulation increases for an initial duration of the flow. This corresponds to the formation stage of the vortex ring. This formation process occurs over an axial distance of $\approx$ $1D$. At the end of the formation stage, all four solutions have the same values of ring circulation. The circulation contained in the shear-layer feeding the vortex ring is thus conserved in all the four solutions. We also see from the plots that, for water, the ring circulation remains nearly a constant during the free shear motion of the vortex ring (after the formation stage ends). However, the polymer solutions show a reduction in ring circulation during free shear motion. Furthermore, the reduction in ring circulation is more drastic as the polymer concentration increases. This reduction in the ring circulation during free shear movement can be explained by the shear thinning viscosities of the polymer solutions. The propagating vortex ring is surrounded by ambient fluid which is at higher viscosity. The zero-shear viscosity $\eta_0$ of, say, the PAMH 25 $ppm$ solution is 20-60 times the viscosity of water. Hence, the moving vortex ring experiences higher viscous dissipation at the edges, leading to a reduction in circulation. This reduction in circulation causes the ring to move slower, thereby reducing the shear rate within the ring, which aids further viscous dissipation due to the increase in viscosity. This mechanism of viscous dissipation of circulation could also explain the degree of reduction of ring circulation of the PAMH 10 $ppm$ and the PAMH 5 $ppm$ solutions - the lower value of $\eta_0$ causes a lesser reduction in ring circulation. The area kinetic energy is the kinetic energy evaluated for the whole field, and is plotted in non-dimensional units in Fig. 3a. It is evident from the plot that, at the end of the formation stage, kinetic energy in the field is nearly the same for all the fluids. However, as the vortex ring propagates further downstream, we observe a reduction in the kinetic energy for PAMH 10 $ppm$ and 25 $ppm$ solutions. As explained for the ring circulation, this reduction in kinetic energy can also be attributed to the considerable increase in viscosity of these two solutions. \\

\begin{figure*}[h]
	\centering
	\includegraphics[width=\textwidth]{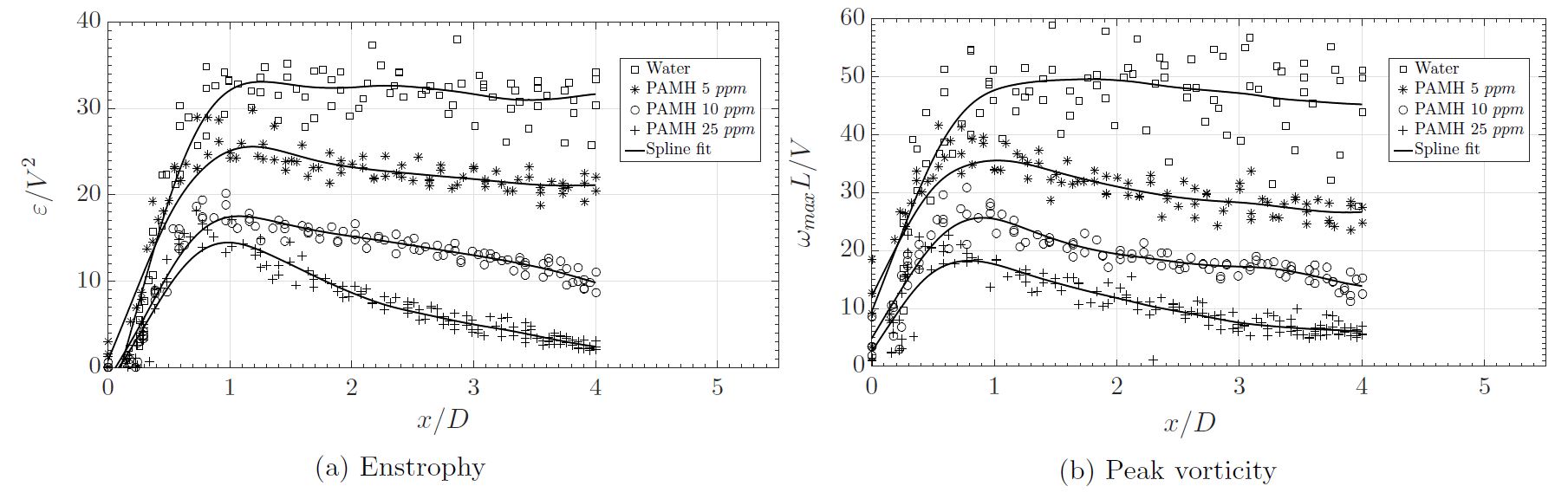}
	\caption{Variation of enstrophy and peak vorticity of the vortex rings for all fluids at L100 V500}
	\label{fig: Peak vorticity and enstrophy constant impulse}
\end{figure*} 

\begin{figure*}[h]
	\centering
	\includegraphics[width=\textwidth]{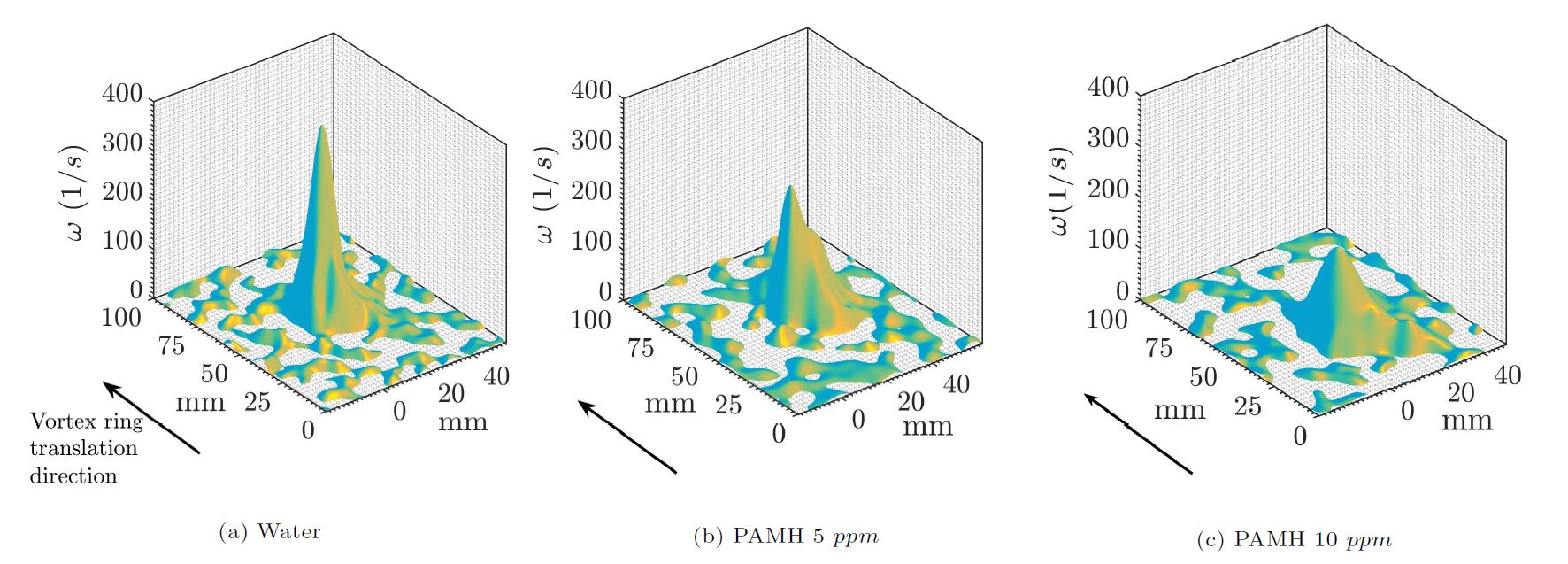}
	\caption{Vorticity distribution over one-half of the vortex ring for Water, PAMH 5 $ppm$ and PAMH 10 $ppm$ at L100 V1000.}
	\label{fig: Vorticity distribution}
\end{figure*}

The variation of peak vorticity and enstrophy are plotted in Fig. \ref{fig: Peak vorticity and enstrophy constant impulse}. We see from the plots that, as the concentration of PAMH increases, the values of enstrophy and peak vorticity reduce during the initial formation stage of the ring. This is a clear difference from the behaviour of the circulation, which showed same values for the different polymer solutions during the formation stage. We present the vorticity distribution on one-half of the ring in Fig. \ref{fig: Vorticity distribution} for water, PAMH 5 ppm and PAMH 10 ppm solutions. It is evident that the polymers in solution modify the vorticity distribution when the impulse is same. Water shows a narrow vorticity distribution with a higher peak, while the polymer solutions show a broader distribution with a lower peak. As the concentration increases, the vorticity distribution becomes flatter. We can now explain the behaviour of the ring circulation, area kinetic energy, enstrophy and peak vorticity as shown in Figs. \ref{fig: Circulation KE constant impulse} and \ref{fig: Peak vorticity and enstrophy constant impulse}: Shear rates in the shear-layer which feeds into vortex ring are $\mathcal{O}(100) 1/s$. Hence, the viscosity encountered by the shear-layer in all the polymer solutions is nearly the same i.e. $\eta_{\infty}$. Because of this, the vorticity (hence circulation) fed to the vortex ring through the shear-layer is conserved, across all the solutions. However, the shear-layer curls up in different manner to form the ring, thereby giving rise to different vorticity distribution in polymer vortex rings. This modification in vorticity distribution is such that, it gets broader with lower peak, as the polymer solution concentration increases. This broadening of the vortex ring core with lower peak values is the reason for reduced enstrophy and peak vorticity at the end of the formation stage.\\ 
% different values of peak vorticity and enstrophy during the formation stage for the different polymer solutions.  We see from figure that, though the circulation is nearly the same in all solutions after the formation stage, peak vorticity and enstrophy differ a lot. More the polymer solution concentration, lesser is the peak vorticity and enstrophy at the end of formation stage. This implies that, vorticity distribution at the end of the formation stage is different in polymer solutions when compared with water. However, the modification of vorticity field is such that circulation is conserved at the end of the formation stage. We speculate that, shear-layer though conserves the vorticity (hence circulation), curls up in different manner to form the ring, thereby giving rise to different vorticity distribution in polymer vortex rings. During free shear motion, both peak vorticity and enstrophy reduces, as is the case with circulation. Polymer solutions with higher concentration exhibit higher rate of reduction. \\
% From figure it is evident that in polymer solutions vorticity distribution is different from that of water even at the largest impulse experiments carried out. Peak vorticity is reduced considerably in polymer solutions. Changes observed in circulation, enstrophy and peak vorticity etc can be explained based on the changes in vorticity distribution. \\

\begin{figure*}[h]
	\centering
	\includegraphics[width=\textwidth]{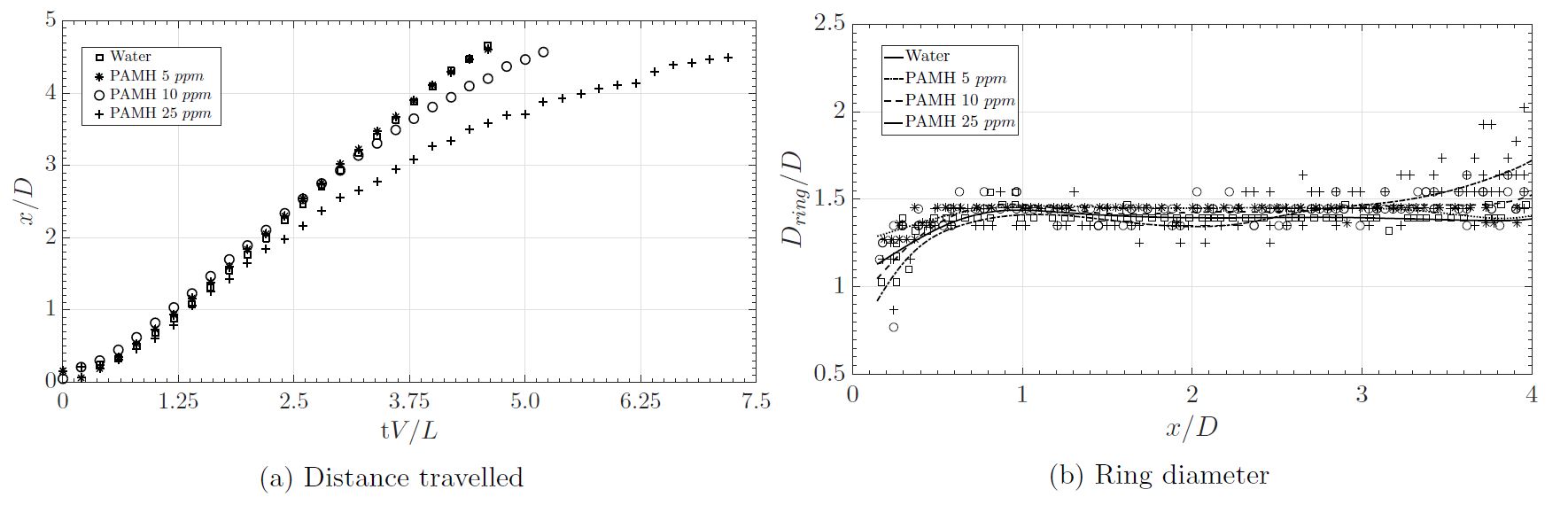}
	\caption{Distance travelled by the vortex ring and ring diameter for all fluids at L100 V500.}
	\label{fig: const_impulse_distance_diameter}
\end{figure*} 

Fig. \ref{fig: const_impulse_distance_diameter} shows the distance travelled by the rings and the ring diameter in non-dimensional units for all fluids at L100 V500. The distance travelled by the vortex ring is a function of translational velocity, which is in turn dependent on the ring circulation, ring diameter and core thickness. We observe that for the first 2$D$ distance, vortex rings in all polymer solutions travel with same speed as that of water rings. We speculate that, for the initial distance, rings are driven by inertia of the fluid pushed by piston, rather than vortex ring's self induced velocity. After the first phase of inertia driven movement, vortex rings travel by their self induced velocity. We see, from the distance traveled plots (Fig. 6a), that the increase in concentration of the PAMH solution increases the deviation of the curve when compared to that of Water. The significant reduction in ring circulation for the PAMH 10 $ppm$ and PAMH 25 $ppm$ solutions during their free shear motion contribute to the deviation from Newtonian behaviour obsevered. The behaviour of the ring diameter is shown in Fig. 6b. The plot shows that, the ring diameter is unchanged in the presence of dissolved polymers in solution for the majority of the ring's motion. However, the PAMH 25 $ppm$ solution shows an enhanced diameter beyond a distance of 3$D$. This increase in diameter is also present in the PAMH 10 $ppm$ solution, but to a lesser extent. The increase in ring diameter is associated with the reduction in circulation and translational velocity during the free shear motion of the ring.\\

We see from the experiments at matched impulse that the vorticity distribution of the vortex ring core is significantly altered by the polymer solutions, and this alteration leads to changes in the peak vorticity and enstrophy in the formation stage of the vortex ring. The circulation, kinetic energy, distance travelled and ring diameter show similar values during the formation stage, but begin to deviate from Newtonian behaviour during the free-shear motion of the ring. In a Newtonian fluid, the normalised vorticity distribution for propagating vortex rings at all times (or axial locations) are self-similar Gaussian profiles \cite{weigand1997evolution}. The polymer solutions clearly break this self-similarity, thereby causing the observed changes in the vortex ring properties. Furthermore, difference in vorticity distribution gets amplified as vortex rings traverse downstream, thereby showing increased deviation in the properties measured for the polymer solutions as compared to water. However, it has to be kept in mind that the Reynolds numbers of water vortex rings and the polymer vortex rings are not matched. Hence, to investigate the possible contribution of this Reynolds number mismatch to the above observations, we have carried out Reynolds number matched experiments, which is explained in the following section. \\

% In conclusion, vorticity distribution over the core of the vortex ring is modified by the addition of polymers. This modification appears as early as during formation stage itself. Though at the end of the formation stage circulation is nearly same for all the solutions used, modification in vorticity distribution is implied from the changes in enstrophy and peak vorticity. This modification is evident, when vorticity over one half of the ring is plotted in 3D, with vorticity as the 3rd axis (As shown in figure \ref{fig: Vorticity distribution}). Modification in vorticity profile can be reasoned by shear-thinning viscosity of polymer solutions. In constant viscosity solution, vorticity evolution will be through a series of self-similar Gaussian profiles. Hence, for all water runs at all time instants, normalized vorticity profiles will be self-similar gaussians \cite{weigand1997evolution}. However, presence of shear thinning viscosity will break this Gaussian self-similarity. \\

% Difference in vorticity distribution gets amplified as vortex rings traverse downstream. Also increasing polymer concentration and reducing initial impulse, exhibit more deviation from Newtonian solution. However, these experiments lack that the Reynolds number of vortex rings involved is not matched. Hence a set of experiments were conducted with $Re$ matched vortex rings, which are presented in next section.

\section{Vortex Rings at Matched Reynolds Numbers}
\label{section: Re matched experiments}

The characteristics of a Newtonian vortex ring are a function of the $L/D$ ratio, piston velocity history and the Reynolds number \cite{shariff1992vortex}. In theory, if these three parameters are matched for vortex rings in two different Newtonian solutions, then further evolution of those rings is same in non-dimensional co-ordinates. In the preceding section, we matched the $L/D$ ratio and the piston velocity history for both Water and the polymer solutions. In this section, we perform experiments by matching the Reynolds numbers of the Water and PAMH vortex rings.\\ 

% Behavior of Newtonian vortex ring mainly depends on three parameters, (1) Reynolds number (2) Ratio of core diameter to ring diameter (3) Normalized vorticity profile. In theory, if these three parameters are matched for vortex rings in two different solutions, then further evolution of those rings is same in non-dimensional co-ordinates. \\

The Reynolds number ($Re = LV/\nu$) is to be matched by increasing $L$ and/or $V$ to compensate for the increase in $\nu$. The PAMH solutions in this study show a shear-thinning behaviour with zero shear and infinite shear plateaus (see Fig. \ref{fig: shear thinning}), which makes defining a single $\nu$ for the matching $Re$ difficult. Multiple methods have been used to calculate the Reynolds number for vortex rings generated in a shear-thinning fluid. Olsthoorn et. al. \cite{olsthoorn2014dynamics} uses $\eta_0$, the zero-shear viscosity, as the reference viscosity to calculate the Reynolds number. Palacios Morales \cite{palacios2013formation} and Palacios Morales et. al. \cite{palacios2015negative} calculate the Reynolds number for a shear thinning fluid using $\eta$ of characteristic shear rate. Bentata et. al. \cite{bentata2018experimental} calculates a generalized Reynolds number, which is obtained by theoretically solving for the friction factor $f$ as a function of $Re$ for a shear thinning fluid in a laminar pipe flow.\\

\begin{figure*}[h]
	\centering
	\includegraphics[width=0.85\textwidth]{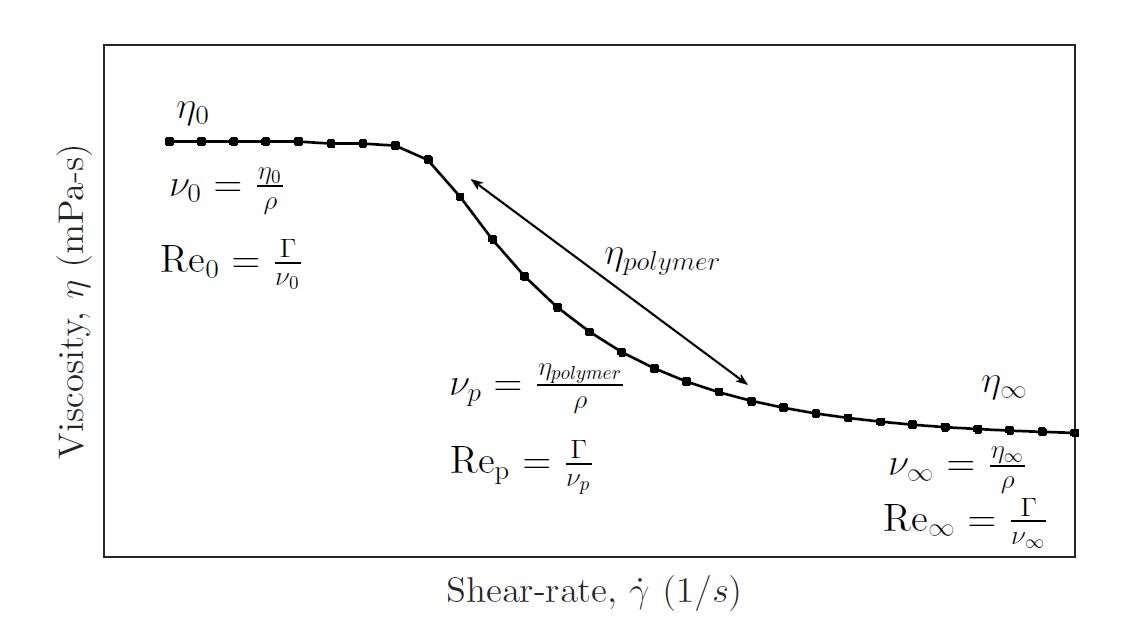}
	\caption{Reynolds number defined for polymer solutions.}
	\label{fig: Re Matching}
\end{figure*}

In this study, we propose a method, which involves enveloping the whole Re range of polymer vortex rings, by considering two limiting Re, instead of matching a particular Re. We define two Reynolds numbers, at the extreme ends of the shear-thinning viscosity profile and conduct water experiments at those Reynolds numbers. By defining two limiting Reynolds numbers, we avoid any misrepresentation of the polymer solutions, since these Reynolds numbers signify the minimum ($Re_0 = V L / \nu_0$) and maximum ($Re_{\infty} = V L / \nu_{\infty}$) values attainable by the polymer solution vortex rings. The actual Reynolds number of the vortex ring in polymer solutions will always lie in between the two limiting values. We show this graphically in Fig. \ref{fig: Re Matching}. The plot shows the shear thinning viscosity profile for a typical polymer solution. The limiting $Re$ values are indicated in the plot.\\

We perform vortex ring experiments at $Re_0$ and $Re_{\infty}$ in water and compare the behaviour of these rings with that of the vortex rings in the polymer solution. We obtain different values in the Reynolds number by varying the circulation of the ring (Note: $\Gamma$ $\sim$ $L$ $\times$ $V$). Specifically, we vary the piston velocity $V$ to vary the circulation and maintain the stroke length $L$ at 100 $mm$ for all the experiments. We keep the stroke length a constant in order to maintain constant values of $L/D$ across all fluids. The values of $\eta_0$ and $\eta_{\infty}$ that are obtained from the Carreau-Yasuda model, as shown in Table \ref{table: Properties of polymer solutions}, are used to calculate $Re_0$ and $Re_{\infty}$. These values of the limiting Reynolds numbers are shown in Table \ref{table: Re Matching parameters} along with all the other parameters of $Re$ matched experiments. For example, the L100 V100 experiments of the PAMH 5 $ppm$ solution will be compared with L100 V86 (corresponding to $Re_0$) and L100 V94 (corresponding to $Re_{\infty}$) of water experiments. We present the experiments at matched Reynolds number for the PAMH 5 $ppm$ and the PAMH 10 $ppm$ solutions.\\

\begin{table*}[h!]
	\begin{tabular}{|c|c|c|c|c|c|}
		\hline
		\multirow{2}{*}{\textbf{\begin{tabular}[c]{@{}c@{}}Polymer\\ solution\end{tabular}}} & \multirow{2}{*}{\textbf{\begin{tabular}[c]{@{}c@{}}$\eta_0$\\ (mPa-s)\end{tabular}}} & \multirow{2}{*}{\textbf{\begin{tabular}[c]{@{}c@{}}$\eta_{\infty}$\\ (mPa-s)\end{tabular}}} & \textbf{In polymer solution}                                & \multicolumn{2}{c|}{\textbf{In water}}                                                                                                                               \\ \cline{4-6} 
		&                                                                                      &                                                                                             & \textbf{\begin{tabular}[c]{@{}c@{}}V\\ (mm/s)\end{tabular}} & \textbf{\begin{tabular}[c]{@{}c@{}}V (To match $Re_0$)\\ (mm/s)\end{tabular}} & \textbf{\begin{tabular}[c]{@{}c@{}}V (To match $Re_{\infty}$)\\ (mm/s)\end{tabular}} \\ \hline
		\multirow{3}{*}{\textbf{5ppm}}                                                       & \multirow{3}{*}{1.05}                                                                & \multirow{3}{*}{0.95}                                                                       & 100                                                         & 86                                                                            & 94                                                                                   \\
		&                                                                                      &                                                                                             & 500                                                         & 429                                                                           & 471                                                                                  \\
		&                                                                                      &                                                                                             & 1000                                                        & 857                                                                           & 943                                                                                  \\ \hline
		\multirow{3}{*}{\textbf{10ppm}}                                                     & \multirow{3}{*}{2.25}                                                                & \multirow{3}{*}{0.955}                                                                      & 100                                                         & 40                                                                            & 94                                                                                   \\
		&                                                                                      &                                                                                             & 500                                                         & 200                                                                           & 471                                                                                  \\
		&                                                                                      &                                                                                             & 1000                                                        & 400                                                                           & 943                                                                                  \\ \hline
 		\multirow{2}{*}{\textbf{25ppm}}                                                      & \multirow{2}{*}{40}                                                                  & \multirow{2}{*}{0.96}                                                                       & 500                                                         & 11.25                                                                         & 471                                                                                  \\
 		&                                                                                      &                                                                                             & 1000                                                        & 22.50                                                                         & 943                                                                                  \\ \hline
	\end{tabular}
	\caption{Input parameters for $Re$ matched experiments. The piston velocity $V$ in water is systematically varied to get the required Reynolds number. The stroke length $L$ is kept constant at 100 $mm$ for all experiments.}
	\label{table: Re Matching parameters}
\end{table*}

As shown in table \ref{table: Re Matching parameters}, L100 V100 experiment of PAMH 5ppm will be compared with L100 V86 and L100 V94 of water experiments. Similarly other experiments in polymer solutions are compared with their respective water counterparts. \\

We compare the matched $Re$ experiments only during the free shear motion of the ring, with the fully formed vortex ring as the starting point (time $t$ = 0 in all the following plots as well as the subscript `0' correspond to the end of the formation phase and the beginning of the free shear motion). We study the effect of matching the Reynolds number on vortex ring properties such as ring circulation, enstrophy, kinetic energy and peak vorticity. We plot these vortex ring properties in non-dimensional units. The vortex ring properties are normalised using their initial values, i.e., the values at the end of the formation stage. These normalised quantities are plotted against non-dimensional time. Time can be non-dimensionalised in two ways: (1) using an inertial time scale $t^* = t \Gamma_0 /{2 \pi R_0^2}$, and (2) using a viscous time scale $t^* = t \nu/{2 \pi R_0^2}$, where, $t$ is dimensional time, $\Gamma_0$ is initial circulation, $R_0$ is initial vortex ring radius and $\nu$ is kinematic viscosity of the solution. Since the vortex rings generated in the current experiments are at high Reynolds numbers ($Re \geq 1000$), the inertial force is more dominant over the viscous force. Hence, the inertial time scale is used for non-dimensionalising time. This method of non-dimensionalising time has been used earlier by Jha and Govardhan \cite{jha2015interaction} and others \cite{asselin2017influence, harris2012instability}.\\

% $Re$ matched experiments are plotted in non-dimensional co-ordinates. Also, $Re$ dependent evolution is only applicable during vortex ring free shear movement, with fully formed vortex ring as the starting point (i.e fully formed vortex ring is considered even at time t = 0). Hence, analysis of vortex formation stage is inherently absent in $Re$ matched experiments. \\

% Properties such as circulation, Enstrophy, Kinetic Energy and Peak vorticity are normalized by their initial values (values at time t=0, i.e values corresponding to vortex ring just after formation stage). There exists two ways of non-dimensionalising time. (1) Using inertial time scale $$t^* = t \times \frac{\Gamma_0}{2 \pi R_0^2}$$ (2) Using viscous time scale $$t^* = t \times \frac{\nu}{2 \pi R_0^2}$$ where, t is dimensional time, $\Gamma_0$ is initial circulation, $R_0$ is initial vortex ring radius and $\nu$ is kinematic viscosity of the solution. Since all our vortex ring experiments are at $Re \geq 1000$, inertial force is the dominant force. Hence, inertial time scale is used for non-dimensionalising, which is earlier used by Jha and Govardhan \cite{jha2015interaction} and others \cite{asselin2017influence, harris2012instability}. \\

We demonstrate the expected behaviour of the circulation of the vortex ring at different Reynolds numbers using a schematic in Fig. 8a. A vortex ring with $Re \rightarrow \infty$ is an inviscid ring, and thus does not encounter any viscous effects. The ring is thus expected to propagate with an unchanged circulation for all time. On the other hand, for $Re \rightarrow 0$, the dominance of viscous forces would cause the vorticity to instantly diffuse, thereby reducing the circulation to zero. The circulation for all finite $Re$ would lie in between these two limits, as shown in the figure.\\

% Variation of circulation as a function of non-dimensional time for limiting value of $Re$ can be intuitively obtained. For $Re \rightarrow \infty$, viscous effects are absent, hence, vorticity won't diffuse, there by $\frac{\Gamma}{\Gamma_0}$ remains constant throughout the duration of the experiment. On the other hand, for $Re \rightarrow 0$, vorticity will diffuse instantly, thereby reducing core circulation to zero. For all finite $Re$, circulation evolution falls in between these two limits, as shown in figure \ref{fig: circulation variation in non-dimensional co-ordinates from theory}. \\

% \begin{figure*}[h!]
% 	\centering
% 	\includegraphics[trim=0 0 0 0,clip,width=0.6\textwidth]{Figure/Reynolds/Circulation/circulation_demo.eps}
% 	\caption{Schematic of expected variation in vortex ring circulation with change in $Re$ in non-dimensional co-ordinates.}
% 	\label{fig: circulation variation in non-dimensional co-ordinates from theory}
% \end{figure*}

\begin{figure*}[h]
	\centering
	\includegraphics[width=\textwidth]{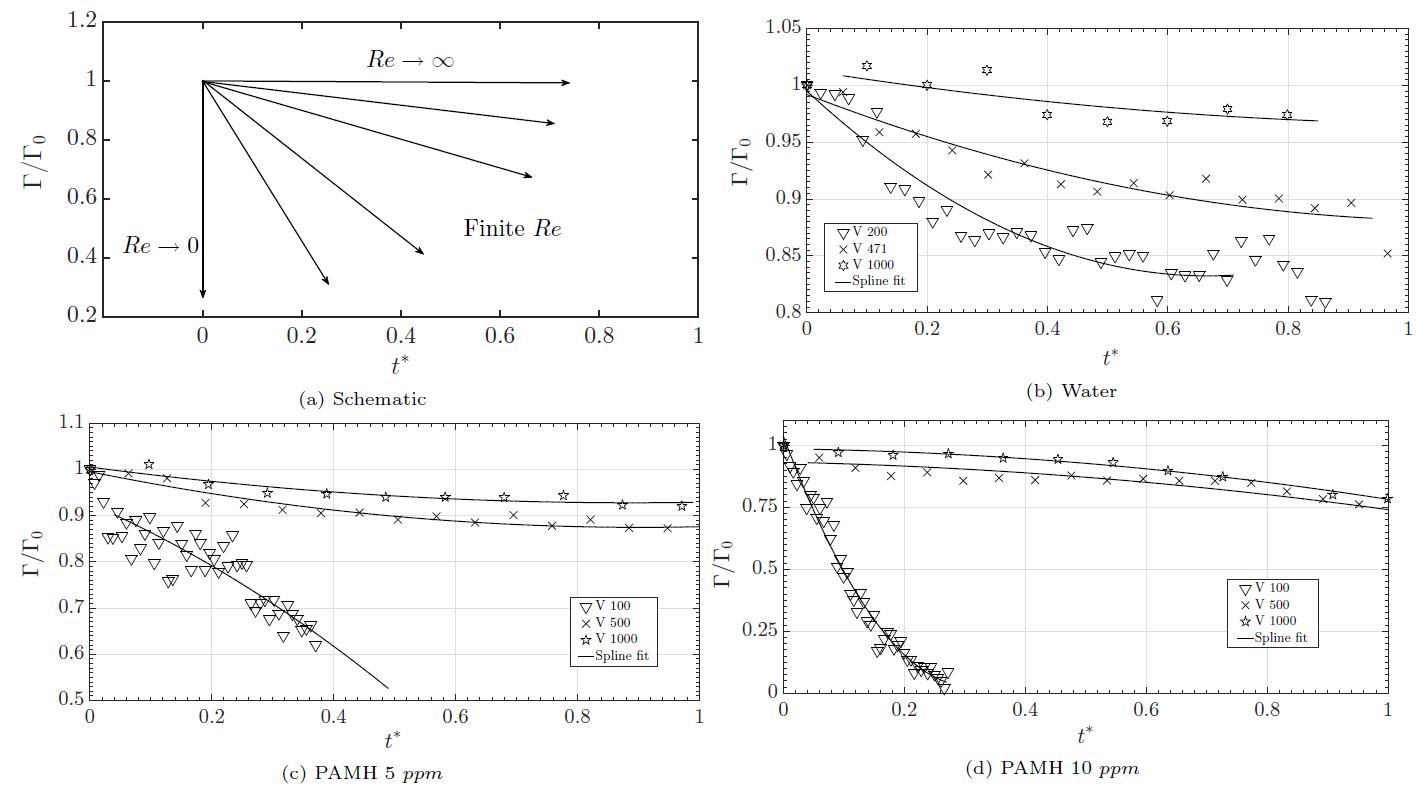}
	\caption{Vortex ring circulation in non-dimensional co-ordinates at different $Re$. (a) shows the expected behaviour of circulation as $Re$ is varied. (b), (c) and (d) show the normalised ring circulation for Water, PAMH 5 $ppm$ and PAMH 10 $ppm$, respectively. Higher $V$ yields a larger $Re$}
	\label{fig: circulation Re matched 5 and 25ppm different runs}
\end{figure*}

We validate the above argument by comparing different $Re$ experiments within each solution. Fig. \ref{fig: circulation Re matched 5 and 25ppm different runs} shows the normalised circulation as a function of non-dimensional time. We clearly observe the transition from an almost constant circulation at high $Re$ to a rapid reduction in circulation as $Re$ reduces. Hence, all the experiments in individual solutions, qualitatively demonstrate the Re based variation. Also, water shows the least variation in the curves of $\Gamma/\Gamma_0$, while the PAMH 10 $ppm$ solution shows the steepest reduction in $\Gamma/\Gamma_0$ at L100 V100. This indicates that, at L100 V100, the PAMH 10 $ppm$ solution has the least $Re$ amongst all the fluids and parameters considered.\\

% Validity of this argument is checked on water and all polymer solutions. As shown in figure \ref{fig: circulation Re matched 5 and 25ppm different runs}, experiments with higher $Re$ show slower reduction in $\frac{\Gamma}{\Gamma_0}$ than the ones with lower $Re$. Hence, all the experiments in individual solutions, qualitatively demonstrate $Re$ based variation. However, the goal is to compare polymer solutions with their respective $Re$ matched water counterparts, and it is shown in figure \ref{fig: Circulation Re matched}. In most cases, $\Gamma^*$ of polymer experiment falls in between that of $Re_0$ and $Re_{\infty}$ experiments, which are conducted in water (Refer figure \ref{fig: Circulation Re matched}). However, in few cases, normalized circulation of polymer solution falls even below that of $Re_0$ after certain time, as shown in figure 12 (b). \\  

\begin{figure*}[h]
	\centering
	\includegraphics[width=\textwidth]{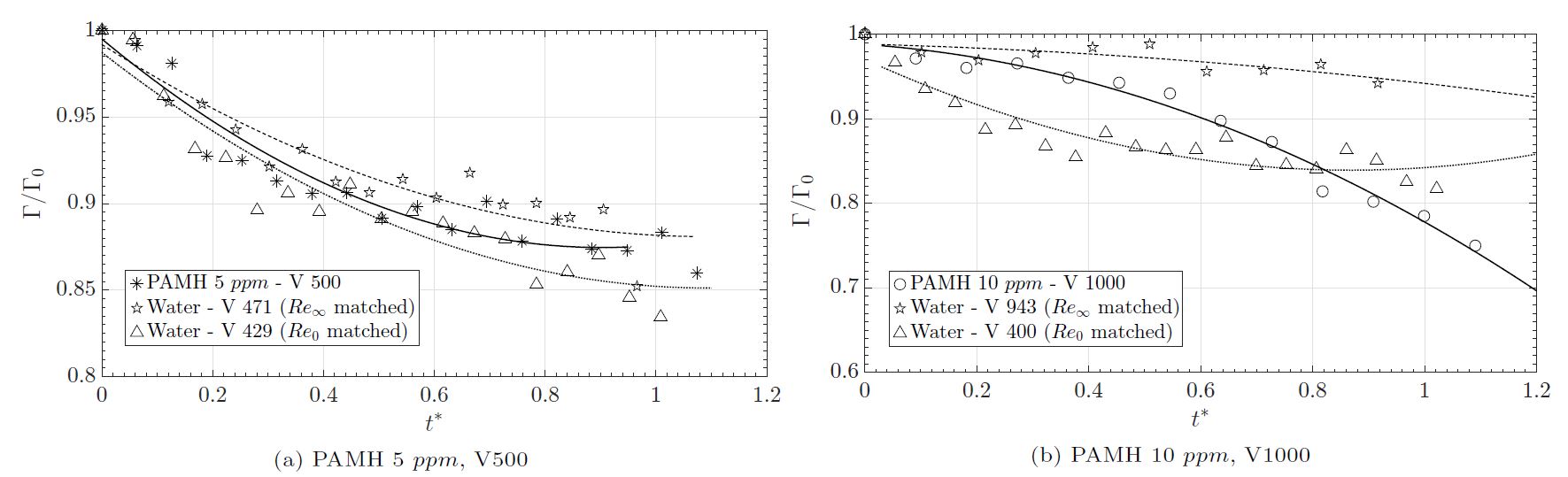}
	\caption{Comparison of normalised vortex ring circulation for polymer solution experiments with its $Re$ matched water experiments. Solid line represent spline fitted data for polymer solution. Dotted lines represent spline fitted data for corresponding water experiments. }
	\label{fig: Circulation Re matched}
\end{figure*}

We now compare the behaviour of polymer solution vortex rings with their respective limiting $Re$ matched water vortex rings. Fig. \ref{fig: Circulation Re matched} shows the normalised ring circulation as a function of non-dimensional time, with Figs. 9a and 9b showing this comparison for the PAMH 5 $ppm$ (at V500) and PAMH 10 $ppm$ (at V1000) solutions, respectively. We observe that $\Gamma/\Gamma_0$ for the PAMH 5 $ppm$ solution lies in between the limiting water curves corresponding to $Re_0$ and $Re_{\infty}$. A similar behaviour is observed for the PAMH 10 $ppm$ solution, but only until a point. Beyond $t^*$ $\approx$ 0.8, the circulation of the polymer ring reduces below the limiting curve of water corresponding to $Re_0$. This indicates that the polymer solutions are initially within the limiting cases prescribed by water, but soon reduce to values below the lowest limit of $Re$.\\

\begin{figure*}[h]
	\centering
	\includegraphics[width=\textwidth]{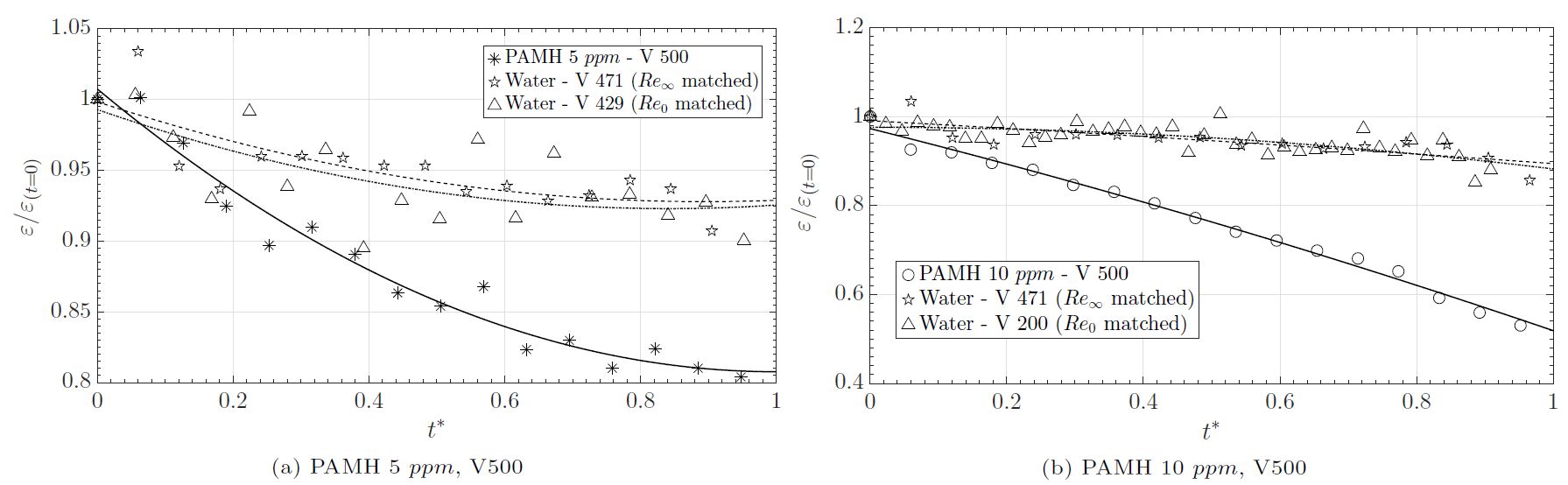}
	\caption{Comparison of vortex ring enstrophy variation for polymer solution experiments with its $Re$ matched water experiments. Solid line represent spline fitted data for polymer solution. Dotted lines represent spline fitted data for corresponding water experiments.}
	\label{fig: Enstrophy Re matched}
\end{figure*}

\begin{figure*}[h]
	\centering
	\includegraphics[width=\textwidth]{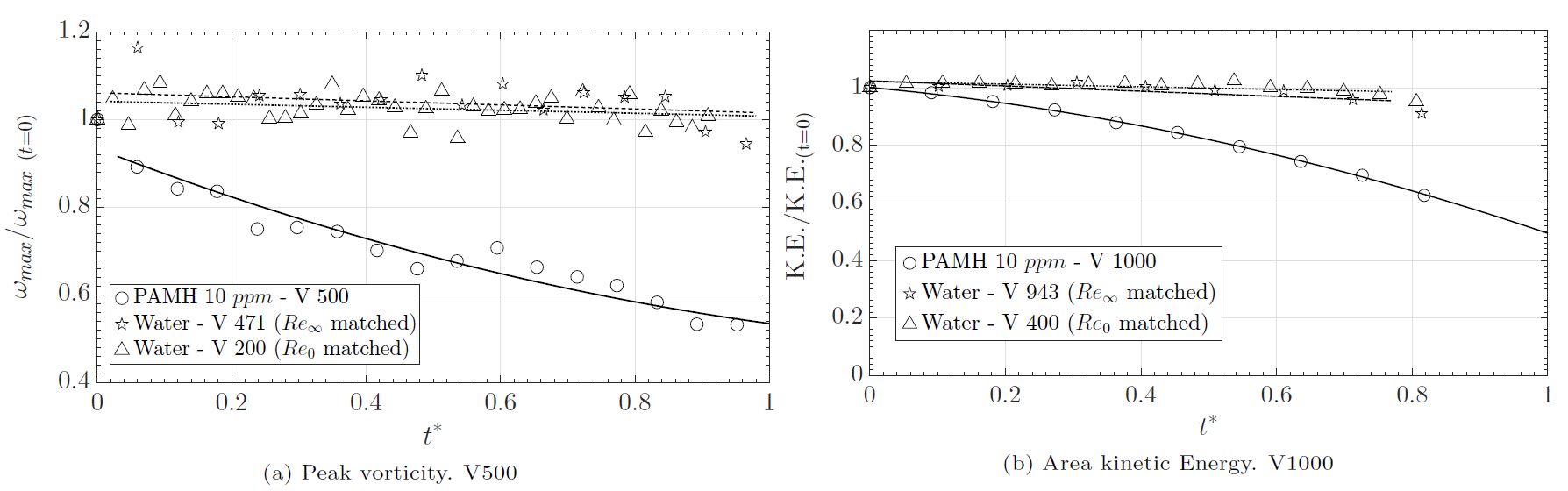}
	\caption{Comparison of peak vorticity and kinetic energy for the PAMH 10 $ppm$ experiment with its $Re$ matched Water experiments. Solid line represent spline fitted data for polymer solution. Dotted lines represent spline fitted data for corresponding water experiments.}
	\label{fig: Peak vorticity Enstrophy Re matched}
\end{figure*}

Fig. \ref{fig: Enstrophy Re matched} shows the variation of the normalised enstrophy for the vortex rings at matched Reynolds number, with Figs. 10a and 10b showing this variation for the PAMH 5 $ppm$ and the PAMH 10 $ppm$ solutions, respectively. We see from the plots that the enstrophy of the PAMH vortex rings are always lesser than that of Water, falling even below the $Re_0$ matched Water vortex ring. This is a clear difference in the behaviour of Newtonian Water and the non-Newtonian PAMH solutions. Fig. \ref{fig: Peak vorticity Enstrophy Re matched} shows the variation of the normalised peak vorticity (Fig. 11a) and the normalised area kinetic energy (Fig. 11b) as a function of non-dimensional time for the PAMH 10 $ppm$ solution. We see, again, that the polymer solution shows a monotonic reduction in both these properties, and shows values considerably lower than the limiting values shown by Water.\\

% From figure \ref{fig: Peak vorticity Enstrophy Re matched}, it is clear that both peak vorticity and kinetic energy reduction is quicker in polymer solution, even when compared with its $Re_0$ matched water experiment. Modification in vorticity distribution is the reason for all the differences, even after matching $Re$. It also points to the inherent limitations of $Re$ matching. Three properties which uniquely describes a vortex ring are $Re$, ratio of core to ring diameter and the vorticity profile. As shown in this section, vorticity profile is different for shear-thinning polymer rings from both $Re_0$ and $Re_{\infty}$ water vortex rings. Hence, just matching $Re$ is not sufficient.\\

\begin{figure*}[h]
	\centering
	\includegraphics[width=\textwidth]{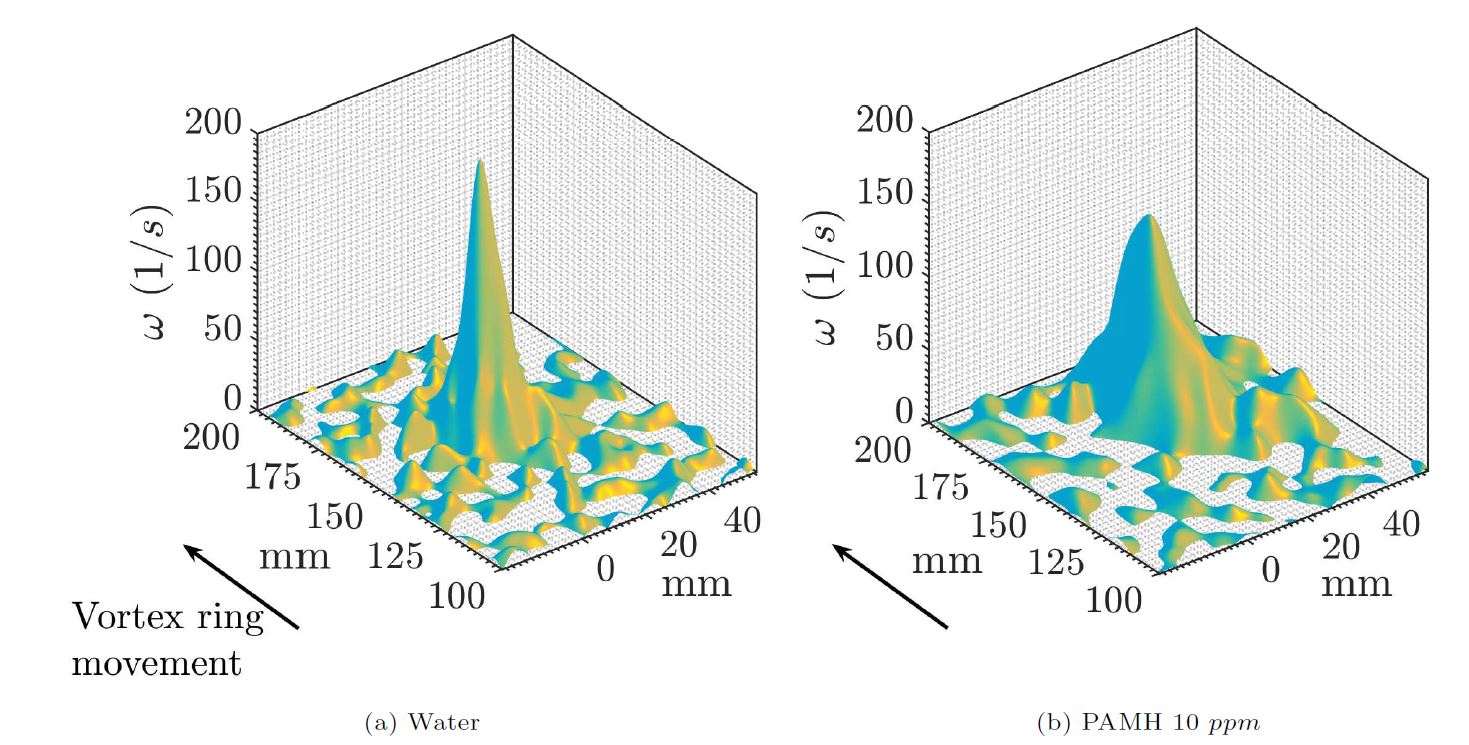}
	\caption{Comparison of vorticity distribution over one half of the ring for PAMH 10 $ppm$ - V 1000 experiment with its $Re_0$ matched water experiment - V400. Vorticity distribution over polymer ring is different even from its $Re_0$ matched experiment in water.}
	\label{fig: enstrophy Re matched comparison 2}
\end{figure*}

We now plot the vorticity distribution over one half of the vortex ring core for the Reynolds number matched experiments. Fig. \ref{fig: enstrophy Re matched comparison 2} shows the vorticity distribution for Water at L100 V400 and the vorticity distribution for the PAMH 10 $ppm$ solution at L100 V1000. Piston parameters of the Water experiment correspond to the lowest Reynolds number $Re_0$ of the PAMH 10 $ppm$ experiment. We see that, vorticity distributions are significantly shorter and broader for the polymer solutions even when compared to its low-limiting $Re_0$ Water counterpart. This causes the reduction in peak vorticity and enstrophy, as was observed in the constant impulse experiments as well.\\ 

% It is evident from figure \ref{fig: Enstrophy Re matched} that, polymer ring looses enstrophy very quickly, though its limiting $Re$ experiments in water doesn't show such a behavior. It is because of stark difference in vorticity distribution in polymer solutions, which is different from water experiments even at limiting $Re$, as shown in figure \ref{fig: enstrophy Re matched comparison 2} (Also refer figure \ref{fig: Vorticity distribution}).\\

In this section, we have studied the effect of polymer solutions on the vortex ring by matching the Reynolds numbers for the Newtonian and the non-Newtonian rings. We find, similar to our observations at constant impulse, that the polymer solutions modify the vorticity distribution in the core of the vortex ring. This leads to lower values of the vortex ring properties, even when compared to the corresponding values at $Re_0$ (which is the lowest $Re$ attainable by the polymer solutions) for Water. We conclude that matching the Reynolds numbers is insufficient to explain the behaviour of the non-Newtonian vortex rings.\\

\section{Formation Number}
\label{section: formation number}
The vortex ring forms when a slug of fluid pushed out of a nozzle rolls up. As the volume of the slug of fluid pushed out by the piston-cylinder mechanism increases, the vortex ring proportionately grows in size and circulation. Gharib et. al. showed \cite{gharib1998universal}, in a seminal work, that the vortex ring cannot grow indefinitely and that there exists a saturation of circulation for the vortex ring. This saturation of ring circulation occurs at $L \approx 4 \times D$. Beyond this value, any additional fluid ejected follows the ring as a trailing jet. This limiting non-dimensional stroke length is called the formation number. \\

The formation number has garnered attention due to its applications in underwater vehicles and bio-inspired propulsion \cite{dabiri2009optimal}. Interestingly, this limiting  $L/D$ ratio is non-existent for vortex pairs, which are the 2-D Cartesian analogue of vortex rings \cite{gao2016formation, afanasyev2006formation, pedrizzetti2010vortex}. Krueger et. al. \cite{krueger2006formation} showed that the formation number changes in the presence of a background co-flow or counter-flow. Dabiri \cite{dabiri2005starting} has studied the effect of temporally varying nozzle exit diameter on vortex ring formation number. Gharib et. al. \cite{gharib2006optimal} demonstrated that cardiac health can be ascertained by measuring the efficiency of the heart in generating optimal vortex rings. This provides a necessary impetus to study the effect of non-Newtonian properties on the formation number of vortex rings \\

In this section, we conduct a set of experiments to study the effect of polymer solutions on the $L/D$ ratio. We generate vortex rings at a single piston velocity ($V$ = 500 $mm/s$) and four different values of stroke lengths corresponding to $L/D$ values of 2.88, 4.32, 5.76 and 7.2 for the PAMH 5 $ppm$, the PAMH 10 $ppm$ and the PAMH 25 $ppm$ solutions. For a Newtonian vortex ring, the trailing jet is  absent at $L/D$ = 2.88, a weak trailing jet is present at $L/D$ = 4.32 and a strong trailing jet is present at $L/D$ = 5.76 and $L/D$ = 7.2.\\  

% A set of experiments were conducted with varying $\frac{\mathrm{L}}{\mathrm{D}}$ ratio in different polymer solutions, to check the effect of non-newtonian properties on vortex ring formation number. Figure \ref{fig: formation_no_5ppm}, \ref{fig: formation_no_10ppm} and \ref{fig: formation_no_25ppm} shows vorticity field for vortex ring experiments in PAMH 5, 10 and 25 ppm experiments with varying $\mathrm{L}/\mathrm{D}$ ratio. In all solutions, piston velocity is kept constant at 500 $mm/s$, as piston stroke-length is varied to get different $\frac{\mathrm{L}}{\mathrm{D}}$ ratios. Vorticity scale used for all the images is given below.

\begin{figure*}[h]
	\centering
	\includegraphics[width=0.75\textwidth]{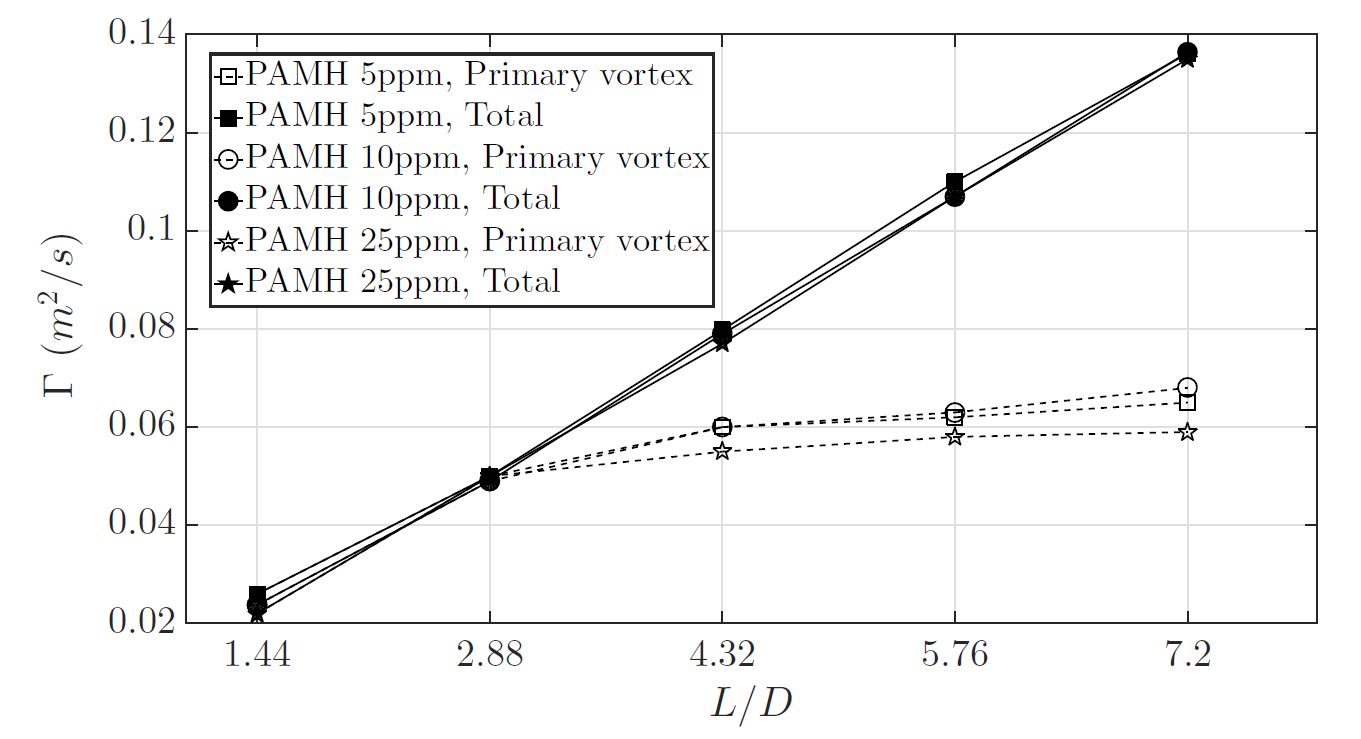}
	\caption{Comparison of primary vortex ring circulation to total circulation}
	\label{fig: Formation number circulation plots}
\end{figure*}

\begin{figure*}[h]
	\centering
	\includegraphics[width=\textwidth]{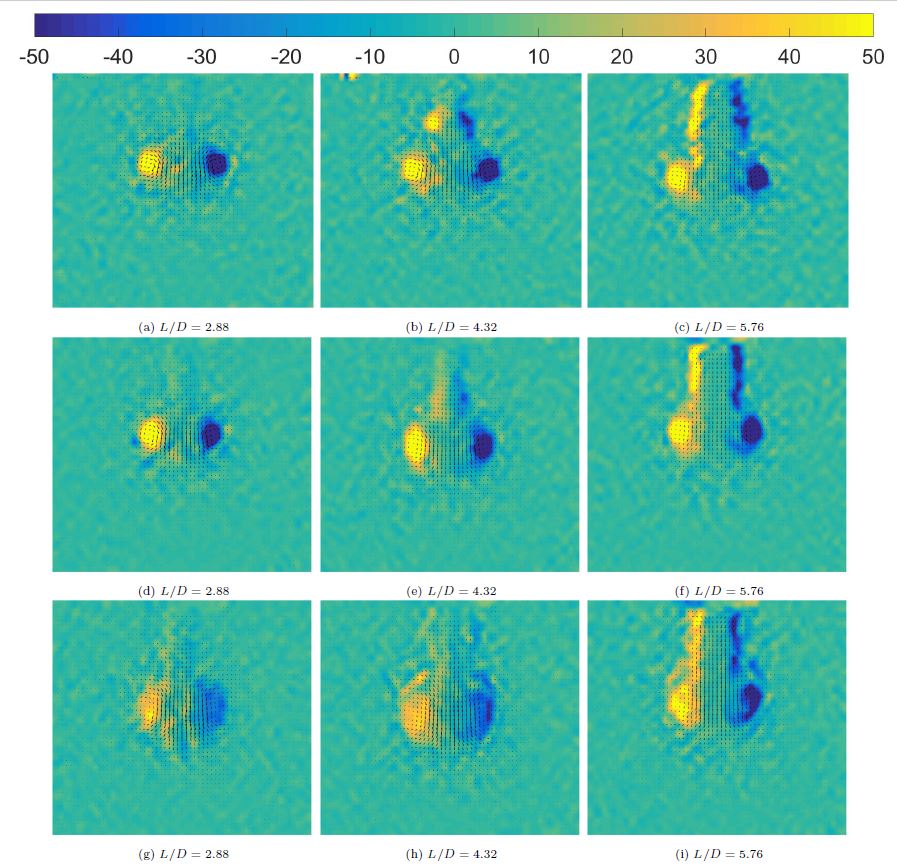}
	\caption{Vorticity field for vortex rings generated in the PAMH solutions at different $L/D$ ratios. (a) -- (c) show vortex rings in the PAMH 5 $ppm$ solution. (d) -- (f) show vortex rings in the PAMH 10 $ppm$ solution. (g) -- (i) show vortex rings in the PAMH 25 $ppm$ solution.}
	\label{fig: formation_no_allfluids}
\end{figure*}

We present the vorticity fields of vortex rings in PAMH solutions in Fig. \ref{fig: formation_no_allfluids} for the first three values of $L/D$ ratio. The behaviour of the PAMH 5 $ppm$ solution can be seen in Figs. 14a to 14c. The vorticity plots clearly show that the vortex ring is the only vortical structure at $L/D$ = 2.88, and no trailing jet is present. At $L/D$ = 4.32, the vortex ring is followed by a trailing jet. At $L/D$ = 5.76, a strong trailing jet is visible behind the vortex ring. The vorticity field of the PAMH 10 $ppm$ solution is shown in Figs. 14d to 14f. We observe, as seen for the PAMH 5 $ppm$ solution, that no trailing jet is present at $L/D$ = 2.88, a weak trailing jet is present at $L/D$ = 4.32 and a strong trailing jet is present at $L/D = 5.76$. The PAMH 25 $ppm$ solution also shows a similar behaviour as the other two PAMH solutions, as seen in Figs. 14g to 14i. Figure \ref{fig: Formation number circulation plots} clearly shows that the primary vortex ring’s circulation saturates, although total circulation increases for $L/D \geq 4$. \\

The vorticity plots seemingly show a trend of a weakening trailing jet as the polymer solution concentration increases. This is merely a consequence of plotting the vorticity fields for the three polymer solutions using a single set of vorticity levels.\\

It is clear from the vorticity maps that the trailing jet is dependent only on the value of the $L/D$ and independent of the polymer concentration. Furthermore, the value of $L/D$ at which the trailing jet begin to form closely correspond to that observed by Gharib et. al. \cite{gharib1998universal}. We find that an isolated vortex ring is formed in the polymer solutions for $L/D \leq 4$, regardless of the concentration. Our observations also agree with those of Palacios-Morales \cite{palacios2013formation}, who noted a formation number of 4 in their vortex ring experiments in shear-thinning fluids.\\

% Formation number of 4 is observed in water, which is in agreement with earlier measurements of \cite{gharib1998universal}. From figure \ref{fig: formation_no_5ppm}, \ref{fig: formation_no_10ppm} and \ref{fig: formation_no_25ppm}, it is clear that there is no difference in formation number for polymer vortex rings also. As long as $\frac{\mathrm{L}}{\mathrm{D}} \leq 4$, an isolated vortex ring is formed in polymer solutions too. A trailing jet is seen only for experiments with $\frac{\mathrm{L}}{\mathrm{D}} \geq 4$. Hence, addition of polymers which is earlier shown to change vorticity distribution, circulation, enstrophy etc, doesn't have any effect on formation number (Note: Structure of vortex rings in PAMH 25ppm solution is clearly different from water rings). Our observations regarding formation number are in perfect agreement with Palacious-Morales and Zenit \cite{palacios2013formation}, who also noted the formation number of 4 for shear-thinning fluids. Vorticity strength of the trailing jet increases with increase in $\frac{\mathrm{L}}{\mathrm{D}}$ ratio. Hence, trailing jet is more clearly observed at higher $\mathrm{L}/\mathrm{D}$ ratios.

\section{Elastic effects}
\label{section: elastic effects}
In this section, we report the behaviour of `ring reversal' that was observed as the vortex rings came to a halt. The reduction in ring circulation causes the ring to continuously decelerate as it propagates forward. As the ring nears the end of its motion, elastic forces start to play a significant role. Palacious-Morales et. al. \cite{palacios2015negative} have reported formation of a `negative' vortex ring ahead of the primary vortex ring. This `negative' vortex ring halts the motion of the primary ring before causing it to disappear altogether. We perform planar laser induced fluorescence (PLIF) visualization to observe the behavior of the ring for PAMH 25 $ppm$ solution.\\

\begin{figure*}[h]
	\centering
	\includegraphics[width=\textwidth]{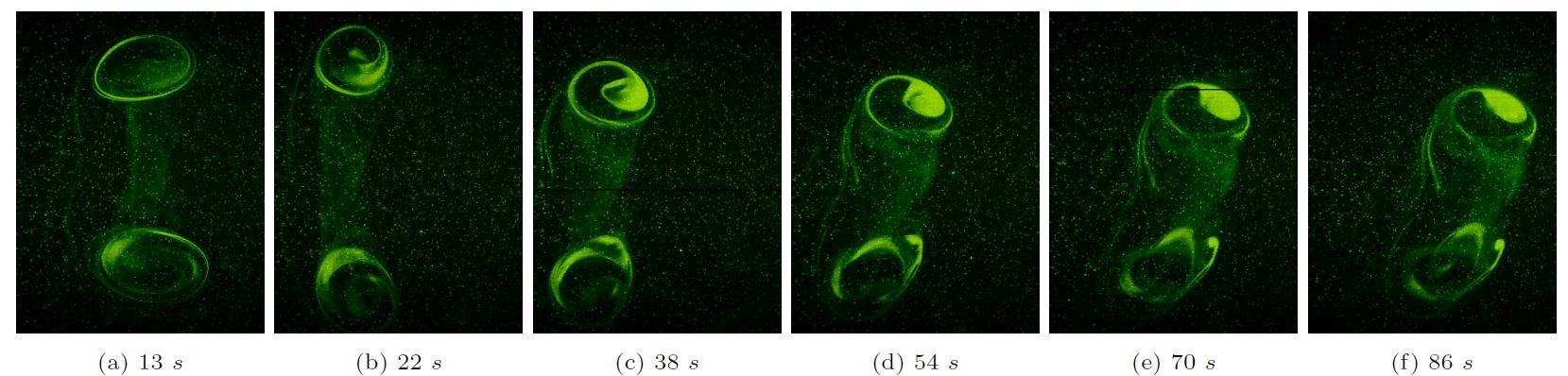}
	\caption{Ring reversal in PAMH 25 $ppm$ vortex rings at L100 V100. The forward motion of the ring is from right to left. The time steps mentioned correspond to the time elapsed since the formation of the vortex ring. The field of view is kept fixed, with the axial width corresponding to $\approx$ 1$D$ (4$D$ -- 5$D$)}
 \label{fig:RingReversalPAMH25Vel100}
\end{figure*} 

\begin{figure*}[h]
	\centering
	\includegraphics[width=\textwidth]{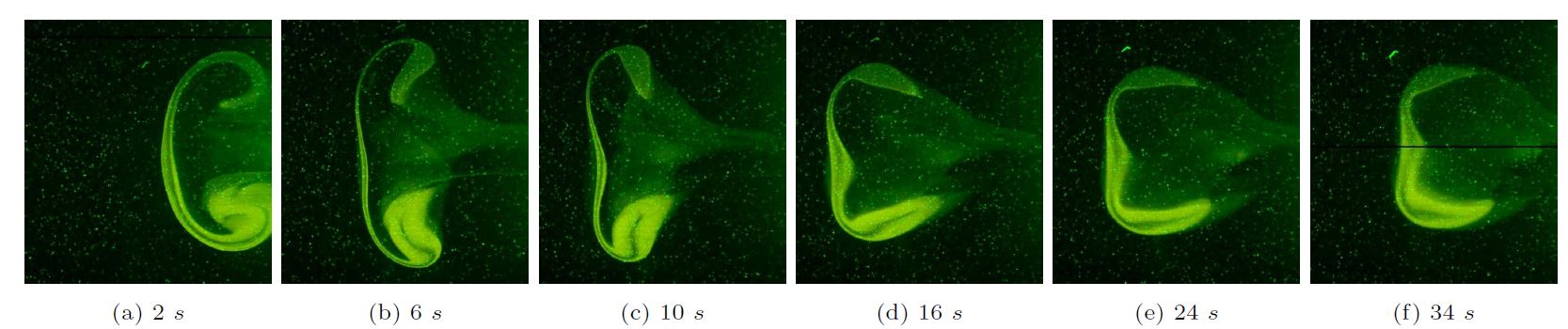}
	\caption{Ring reversal in PAMH 25 $ppm$ vortex rings at L50 V50. The forward motion of the ring is from right to left. The time steps mentioned correspond to the time elapsed since the formation of the vortex ring. The field of view is $\approx$ 1$D$ (1$D$ -- 2$D$)}
 \label{fig:RingReversalPAMH25Vel50}
\end{figure*} 

Fig. \ref{fig:RingReversalPAMH25Vel100} shows the `ring reversal' phenomenon in the PAMH 25 $ppm$ solution at L100 V100. The sequence of images depict the change in direction of the ring's motion. The forward motion of the ring ends in Fig. \ref{fig:RingReversalVel100Img775}. The ring then starts `unrolling' itself by moving in the opposite direction. This process is a slow one, as indicated by the timestamps. This `unrolling' process is better observed at a lower impulse of L50 V50, as seen in Fig. \ref{fig:RingReversalPAMH25Vel50}. The forward motion of the vortex ring is quickly retarded, with the ring being pulled back towards the nozzle. An interesting behaviour observed here is the tendency of the polymer solution to prevent the ring from rolling up - the ring is allowed to roll up to about half a turn (Fig. 16b). The ring is then forced to `unroll' itself. The vorticity levels involved with negative vortex ring are very small and could not be accurately measured in our experiments. \\

The reduction in ring circulation and other vortex ring parameters have been attributed to the shear thinning viscosity profile \cite{palacios2013formation, bentata2018experimental}. Palacios-Morales et. al. \cite{palacios2015negative} show that elastic effects reinforce shear thinning effects with respect to the size and circulation of the vortex ring. The change in the vortex ring properties could be due to a combination of shear thinning effects and elastic effects with their individual contributions being unknown. It is thus difficult to decouple the effects of shear thinning and elasticity. The `ring reversal' phenomenon is clearly an elastic phenomenon. The lack of ring inertia causes the ring to experience low shear rates, which corresponds to the zero shear plateau. The effect of shear thinning is thus non-existent. \\

% As it is reasoned earlier, shear-thinning effects can disrupt Gaussian self-similar vorticity distribution. Hence, shear-thinning alone can cause the change in vorticity distribution and hence, all the changes observed in section \ref{section: Matched Impulse experiments, section: Re matched experiments}. However, Palacious-Morales et al \cite{palacios2015negative} have shown that elastic effects reinforce shear-thinning effects as far as size and circulation of the ring is considered, thus making it very hard to separate the two effects. Hence, it has to be kept in mind that elastic effects could still have some smaller contribution in the changes observed during formation and free-shear motion of the ring, although shear-thinning is the dominant factor. However, during the dissipation of the vortex ring, elastic effects show a clear prominence, mainly because of two factors. Small inertial force involved is the first factor. Second factor is that whole flow field has nearly $\eta_0$ viscosity, hence, shear-thinning behavior ceases to exist. More careful studies along with improved rheological measurements are necessary to clearly probe the elastic effects and to separate it from shear-thinning effect.
     
\section{Conclusion}
\label{section: conclusion}

We have conducted experiments on vortex rings in non-Newtonian polymer solutions. We have studied the effect of these polymer solutions on vortex ring properties such as ring circulation, enstrophy, kinetic energy, etc. We observe that the vortex ring parameters are considerably altered in the presence of the polymer solutions. Vortex ring properties such as the enstrophy and peak vorticity show significant differences when compared to Newtonian Water during the initial formation process of the ring. The kinetic energy and circulation are unaffected in this stage due to the conservation of the product of vorticity and area of the vorticity distribution in the polymer solutions. Once the vortex ring completes the formation process, all the vortex ring properties are severly affected by the polymer solutions.\\   

The vortex rings generated in the polymer solutions were compared with Newtonian water in two ways: (1) by matching impulse generated by the piston, and (2) by matching the Reynolds number. The Reynolds number was matched at viscosities corresponding to the zero shear and the infinite shear plateaus polymer solution. The modification of the vorticity distribution was evident in both these sets of experiments. It is interesting to note that the vorticity distribution of the polymer solution was significanly modified, even when compared to the vorticity distribution for Water at a Reynolds number corresponding to $\eta_0$. We therefore suggest that it is this modification of the vorticity distribution that causes the changes in the vortex ring paramters observed. Some parameters, such as the ring circulation, show the effects of the modified vorticity distribution only after some time has elapsed, whereas most of the parameters, such as enstrophy and peak vorticity, show an immediate effect in their values.\\

We have studied the behaviour of vortex rings in polymer solutions with water-like viscosities and at high Reynolds numbers. These parameters are relevant to turbulent drag reduction, and the application of that phenomenon to industrial applications such as oil pipelines. We find that the behaviour of the vortex rings observed in this study is similar to that observed in previous work by Palacios-Morales \cite{palacios2013formation} and Bentata et. al. \cite{bentata2018experimental}, whose vortex rings were at significantly lower Reynolds numbers. We also demonstrated the `ring reversal' phenomenon that is observed at low inertia situation of the vortex ring when the elastic effects are dominant. This behaviour qualitatively agrees with the `negative vortex ring' that was observed by Palacios-Morales et. al. \cite{palacios2015negative}. We also explored the effect of polymer solutions on the formation number of the vortex ring. We don't observe any changes in the formation number for the vortex rings in polymer solution. The exact contribution of the shear thinning effects and viscoelastic effects on the vortex ring properties observed in this study is still not known. Experiments which can decouple these effects, particularly at high Reynolds numbers would help in understanding this interaction of dissolved polymers in solution with canonical vortex structures such as the vortex ring. Such experiments might also help in better understanding the complicated phenomenon of turbulent polymer drag reduction.\\

\bibliography{references}
\end{document}